%% file: main.tex
\documentclass[lettersize,journal]{IEEEtran}
\IEEEoverridecommandlockouts

\input{Sections/0_preamble}
\input{Sections/0_acronym}

\begin{document}

\title{Attention in Motion: Secure Platooning via Transformer-based Misbehavior Detection}

\author{
    Konstantinos Kalogiannis*\orcidicon{0000-0002-4656-2565}%
    ,~\IEEEmembership{Student Member,~IEEE,}
    Ahmed Mohamed Hussain*\orcidicon{0000-0003-4732-9543}%
    ,~\IEEEmembership{Member,~IEEE,}
    Hexu Li\orcidicon{0009-0004-2430-1374}
    , and Panos Papadimitratos\orcidicon{0000-0002-3267-5374}%
    ,~\IEEEmembership{Fellow,~IEEE}%
    \thanks{*Equally contributing authors. Konstantinos Kalogiannis, Ahmed Mohamed Hussain, and Panos Papadimitratos are with the Networked Systems Security (NSS) Group, KTH Royal Institute of Technology, Stockholm, Sweden. Email: \{konkal, ahmhus, papadim\}@kth.se. Hexu Li is with the Tech Validation – AI department at the Chief Technology Officer (CTO) Organization, Lenovo, Beijing, China. Email: lihx36@lenovo.com.
    }%
}

\markboth{IEEE Transactions on Intelligent Transportation Systems}%
{\MakeLowercase{\textit{Kalogiannis et al.}}:Attention in Motion: Secure Platooning via Transformer-based Misbehavior Detection}

\maketitle

\begin{abstract}
Vehicular platooning promises transformative improvements in transportation efficiency and safety through the coordination of multi-vehicle formations enabled by Vehicle-to-Everything (V2X) communication. However, the distributed nature of platoon coordination creates security vulnerabilities, allowing authenticated vehicles to inject falsified kinematic data, compromise operational stability, and pose a threat to passenger safety. Traditional misbehaviour detection approaches, which rely on plausibility checks and statistical methods, suffer from high False Positive (FP) rates and cannot capture the complex temporal dependencies inherent in multi-vehicle coordination dynamics. We present \underline{A}ttention \underline{I}n \underline{M}otion (\textsc{AIMformer}), a trans\underline{former}-based framework specifically tailored for real-time misbehaviour detection in vehicular platoons with edge deployment capabilities. 
\rev{\textsc{AIMformer} leverages multi-head self-attention mechanisms to capture intra-vehicle temporal dynamics, with a spatio-temporal variant that further models inter-vehicle spatial correlations. It incorporates global positional encoding with vehicle-specific temporal offsets to handle join/exit maneuvers.}
We propose a Precision-Focused Binary Cross-Entropy (PFBCE) loss function that penalizes FPs to meet the requirements of safety-critical vehicular systems. Extensive evaluation across 4 platoon controllers, multiple attack vectors, and diverse mobility scenarios demonstrates superior performance ($\geq$ 0.93) compared to state-of-the-art baseline architectures. A comprehensive deployment analysis utilizing TensorFlow Lite (TFLite), Open Neural Network Exchange (ONNX), and TensorRT achieves sub-millisecond inference latency, making it suitable for real-time operation on resource-constrained edge platforms. Hence, validating \textsc{AIMformer} is viable for both in-vehicle and roadside deployment. 

\end{abstract}


\input{Sections/1_introduction}

\input{Sections/2_background}

\input{Sections/3_adversary_system}
\input{Sections/4_implementaion}
\input{Sections/5_perfromance_new}

\input{Sections/6_discussion}

\input{Sections/7_related_work}

\input{Sections/8_conclusion}

\balance
\bibliographystyle{IEEEtran}
\bibliography{main}

\vspace{-33pt}

\begin{IEEEbiographynophoto}{Konstantinos Kalogiannis} is a PhD Candidate in the Networked Systems Security (NSS) Group at KTH Royal Institute of Technology, Stockholm, Sweden. His research interests include security and privacy in cyber-physical systems, with a focus on vehicular ad hoc networks (VANETs) and platooning, and the security analysis of large-scale software systems.
   
\end{IEEEbiographynophoto}
\vspace{-33pt}
\begin{IEEEbiographynophoto}{Ahmed Mohamed Hussain}
is currently a Doctoral Researcher with the Networked Systems Security (NSS) Group at KTH Royal Institute of Technology, Stockholm, Sweden. His research interests encompass IoT security and privacy, wireless security, Edge AI, and the application of artificial intelligence to cybersecurity challenges. His homepage: https://www.ahmed-hussain.net
\end{IEEEbiographynophoto}
\vspace{-33pt}
\begin{IEEEbiographynophoto}{Hexu Li} is currently a Researcher working with the Tech Validation – AI department at the Chief Technology Officer (CTO) Organization, Lenovo, Beijing, China. He received his Master's degree in Communication Systems from KTH Royal Institute of Technology, Stockholm, Sweden. 
\end{IEEEbiographynophoto}
\vspace{-33pt}
\begin{IEEEbiographynophoto}{Panos Papadimitratos} earned his Ph.D. degree from Cornell University, Ithaca, NY, USA. He leads the Networked Systems Security group at KTH, Stockholm, Sweden. Panos is an IEEE Fellow, a Young Academy of Europe Fellow, a Knut and Alice Wallenberg Academy Fellow, and an ACM Distinguished Member. His group webpage is: www.eecs.kth.se/nss.
\end{IEEEbiographynophoto}

\vfill

\end{document}

%% file: Sections/0_preamble.tex
\usepackage{msc}
\usepackage{algorithmic}
\usepackage{graphicx}
\usepackage{acronym, comment, subcaption}
\usepackage{tabularx, pifont}
\usepackage[ruled,vlined,linesnumbered]{algorithm2e}
\usepackage{hyperref} 
\usepackage{mathtools}
\usepackage{pdflscape}
\usepackage{lipsum}
\usepackage{booktabs}
\usepackage{multirow}
\usepackage{balance}
\usepackage{amssymb}
\usepackage{amsfonts}
\usepackage{booktabs}
\usepackage{float}
\usepackage{subcaption}
\usepackage{svg}
\usepackage{enumitem}
\usepackage{subcaption}
\usepackage{colortbl}
\usepackage{xcolor}
\usepackage{etoolbox}
\usepackage{lscape}

\usepackage{wasysym}

\usepackage{pgfplots}
\pgfplotsset{compat=1.18}

\usepackage{tikz}
\usetikzlibrary{chains,shapes.multipart}
\usetikzlibrary{shapes,calc}
\usetikzlibrary{automata,positioning}

\newcommand{\cmark}{\ding{51}}
\newcommand{\xmark}{\ding{55}}

\graphicspath{{Figures/}}

\usepackage[most]{tcolorbox}

\definecolor{bbyblue}{HTML}{89cff0}
\definecolor{gblue}{HTML}{4f79a7}
\definecolor{ggreen}{HTML}{77b7b2}
\definecolor{gred}{HTML}{e1575a}
\definecolor{gorange}{HTML}{f28e2a}
\definecolor{ppurple}{HTML}{603a70}
\definecolor{mydarkblue}{rgb}{0,0.08,0.45}

\tcbset{
  takeaway/.style={
    width=\linewidth,
    top=8pt,
    bottom=4pt,
    colback=ppurple!6!white,
    colframe=ppurple,
    colbacktitle=ppurple,
    enhanced,
    center,
    attach boxed title to top left={yshift=-0.1in,xshift=0.15in},
    boxed title style={boxrule=0pt,colframe=white,},
  }
}

\tcbset{
  aibox/.style={
    width=\linewidth,
    top=8pt,
    bottom=4pt,
    colback=bbyblue!6!white,
    colframe=bbyblue,
    colbacktitle=bbyblue,
    coltitle=black,
    enhanced,
    center,
    attach boxed title to top left={yshift=-0.1in,xshift=0.15in},
    boxed title style={boxrule=0pt,colframe=white,},
  }
}

\newtcolorbox{AIbox}[2][]{aibox,title=#2,#1}
\newtcolorbox{Takeaway}[2][]{takeaway,title=#2,#1}

\usepackage{scalerel}
\usepackage{tikz}
\usetikzlibrary{svg.path}

\definecolor{orcidlogocol}{HTML}{A6CE39}
\tikzset{
  orcidlogo/.pic={
    \fill[orcidlogocol] svg{M256,128c0,70.7-57.3,128-128,128C57.3,256,0,198.7,0,128C0,57.3,57.3,0,128,0C198.7,0,256,57.3,256,128z};
    \fill[white] svg{M86.3,186.2H70.9V79.1h15.4v48.4V186.2z}
                 svg{M108.9,79.1h41.6c39.6,0,57,28.3,57,53.6c0,27.5-21.5,53.6-56.8,53.6h-41.8V79.1z M124.3,172.4h24.5c34.9,0,42.9-26.5,42.9-39.7c0-21.5-13.7-39.7-43.7-39.7h-23.7V172.4z}
                 svg{M88.7,56.8c0,5.5-4.5,10.1-10.1,10.1c-5.6,0-10.1-4.6-10.1-10.1c0-5.6,4.5-10.1,10.1-10.1C84.2,46.7,88.7,51.3,88.7,56.8z};
  }
}

\newcommand\orcidicon[1]{\href{https://orcid.org/#1}{\mbox{\scalerel*{
\begin{tikzpicture}[yscale=-1,transform shape]
\pic{orcidlogo};
\end{tikzpicture}
}{|}}}}
\usepackage{hyperref}

\graphicspath{{Figures/}}

\definecolor{revblue}{RGB}{0,0,0}
\newcommand{\rev}[1]{{\color{revblue}#1}}
\definecolor{revorange}{RGB}{230,120,0}

\usepackage{tikz}
\newcommand{\blackcircled}[1]{\tikz[baseline=(char.base)]{
    \node[shape=circle,draw,fill=black,inner sep=1pt] (char) {\textcolor{white}{#1}};}}

    {\endgroup}

\usepackage{amssymb}

%% file: Sections/0_acronym.tex
\acrodef{WLAN}{Wireless Local Area Network}
\acrodef{UN}{United Nations}
\acrodef{SDG}{Sustainable Development Goal}

\acrodef{ML}{Machine Learning}
\acrodef{SVM}{Support Vector Machine}
\acrodef{DT}{Decision Tree}
\acrodef{RF}{Random Forest}
\acrodef{BSM}{Basic Safety Message}
\acrodef{BiLSTM}{Bidirectional LSTM}
\acrodef{DL}{Deep Learning}
\acrodef{RNN}{Recurrent Neural Network}
\acrodef{LSTM}{Long Short-Term Memory}
\acrodef{CNN}{Convolutional Neural Network}
\acrodef{DDoS}{Distributed Denial-of-Service}
\acrodef{V2X}{Vehicle-to-Everything}
\acrodef{k-NN}{k-Nearest Neighbor}
\acrodef{KDE}{Kernel Density Estimation}
\acrodef{LLM}{Large Language Model}
\acrodef{MDS}{Misbehavior Detection System}
\acrodef{RSS}{Received Signal Strength}
\acrodef{VANET}{Vehicular Ad Hoc Network}
\acrodef{WEKA}{Waikato Environment for Knowledge Analysis}
\acrodef{DS}{Dempster-Shafer}
\acrodef{EL}{Ensemble Learning}
\acrodef{IDS}{Intrusion Detection System}
\acrodef{FLDS}{False Location Detection System}
\acrodef{FFN}{FeedForward Network}
\acrodef{Ens.RF}{Ensemble Random Forest}
\acrodef{CNB}{Complement Naive Bayes}
\acrodef{DTC}{Decision Tree Classifier}
\acrodef{GBC}{Gradient Boosting Classifier}
\acrodef{MDF}{Misbehavior Detection Framework}
\acrodef{MLP}{Multi-Layer Perceptron}
\acrodef{FL}{Federated Learning}
\acrodef{RL}{Reinforcement Learning}
\acrodef{DML}{Deep Multimodal Learning}
\acrodef{IoV}{Internet of Vehicles}
\acrodef{GRU}{Gated Recurrent Unit}
\acrodef{GTU}{Gated Temporal Unit}
\acrodef{BM}{Bookmaker Informedness}
\acrodef{MHA}{Multi-Head Attention}
\acrodef{MK}{Markedness}
\acrodef{MCC}{Matthews Correlation Coefficient}
\acrodef{AI}{Artificial Intelligence}
\acrodef{TFLite}{TensorFlow Lite}
\acrodef{TinyML}{Tiny Machine Learning}
\acrodef{ONNX}{Open Neural Network Exchange}
\acrodef{BCE}{Binary Cross-Entropy}
\acrodef{AAA}{Authentication, Authorization and Accounting}
\acrodef{ACC}{Adaptive Cruise Control}
\acrodef{ACL}{Access Control List}
\acrodef{AD}{Automated Driving}
\acrodef{ADAM}{Adaptive Moment Estimation}
\acrodef{ADAS}{Advanced Driver Assistance Systems}
\acrodef{AKI}{Accountable Key Infrastructure}
\acrodef{API}{Application Programming Interface}
\acrodef{ASS}{Anonymity Set Size}
\acrodef{ATI}{Attack Time Impact}
\acrodef{AUC}{Area Under the Curve}
\acrodef{BSM}{Basic Safety Message}
\acrodef{BIC}{Bayesian Information Criterion}
\acrodef{BYOD}{Bring Your Own Device}
\acrodef{C2C-CC}{Car2Car Communication Consortium}
\acrodef{C2C}{Car-to-Car}
\acrodef{C2I}{Car-to-Infrastructure}
\acrodef{CA}{Certification Authority}
\acrodef{CAN}{Controller Area Network}
\acrodef{CHMM}{Continuous Hidden Markov Model}
\acrodef{CI}{Confidence Interval}
\acrodef{CMIX}{Cryptographic Mix-Zone}
\acrodef{CN}{Common Name}
\acrodef{CVS}{Constant Vehicle Spacing}
\acrodef{CACC}{Cooperative Adaptive Cruise Control}
\acrodef{CAM}{Cooperative Awareness Message}
\acrodef{CAMP VSC3}{Crash Avoidance Metrics Partnership Vehicle Safety Consortium}
\acrodef{CAV}{Cooperative Aware Vehicle}
\acrodef{CDG}{Constant-Distance Gap}
\acrodef{CIA}{Confidentiality, Integrity and Availability}
\acrodef{CPU}{Central Processing Unit}
\acrodef{CRL}{Certificate Revocation List}
\acrodef{CTG}{Constant-Time Gap}
\acrodef{CTH}{Constant Time Headway}
\acrodef{CDN}{Content Delivery Network}
\acrodef{C-IT}{Cooperative Intelligent Transport System}
\acrodef{COCA}{Cornell OnLine Certification Authority}
\acrodef{CSR}{Certificate Signing Request}
\acrodef{DAA}{Direct Anonymous Attestation}
\acrodef{DDoS}{Distributed Denial of Service}
\acrodef{DDH}{Decisional Diffie-Helman}
\acrodef{DENM}{Decentralized Environmental Notification Message}
\acrodef{DHMM}{Discrete Hidden Markov Model}
\acrodef{DHT}{Distributed Hash Table}
\acrodef{DL/ECIES}{Discrete Logarithm and Elliptic Curve Integrated Encryption Scheme}
\acrodef{DNN}{Deep Neural Network}
\acrodef{DoS}{Denial of Service}
\acrodef{DoT}{Department of Transportation}
\acrodef{DPA}{Data Protection Agency}
\acrodef{DSRC}{Dedicated Short Range Communication}
\acrodef{DSS}{Digital Signature Standard}
\acrodef{ECU}{Electronic Control Unit}
\acrodef{Edge AI}{Edge Artificial Intelligence}
\acrodef{EDR}{Event Data Recorder}
\acrodef{ETSI}{European Telecommunications Standards Institute}
\acrodef{ECDSA}{Elliptic Curve Digital Signature Algorithm}
\acrodef{ECC}{Elliptic Curve Cryptography}
\acrodef{EVITA}{E-safety Vehicle Intrusion protected Applications}
\acrodef{FOT}{Field Operational Test}
\acrodef{FPGA}{Field-Programmable Gate Array}
\acrodef{FNR}{False Negative Rate}
\acrodef{FPR}{False Positive Rate}
\acrodef{GPA}{Global Passive Adversary}
\acrodef{GPS}{Global Positioning System}
\acrodef{GN}{GeoNetworking}
\acrodef{GS-VLR}{Group Signatures with Verifier Local Revocation}
\acrodef{GS}{Group Signatures}
\acrodef{GM}{Group Manager}
\acrodef{GBA}{Generic Bootstrapping Architecture}
\acrodef{GNSS}{Global Navigation Satellite System}
\acrodef{GUI}{Graphic User Interface}
\acrodef{HMM}{Hidden Markov Model}
\acrodef{GMM}{Gaussian Mixture Model}
\acrodef{GMMHMM}{Gaussian Mixture Model Hidden Markov Model}
\acrodef{HSM}{Hardware Security Module}
\acrodef{HTTP}{Hypertext Transfer Protocol}
\acrodef{IEEE}{Institute of Electrical and Electronics Engineers}
\acrodef{IETF}{Internet Engineering Task Force}
\acrodef{IoT}{Internet of Things}
\acrodef{ITS}{Intelligent Transport Systems}
\acrodef{IT}{Information Technologies}
\acrodef{IVN}{In-Vehicle Network}
\acrodef{IMSI}{International Mobile Subscriber Identity}
\acrodef{IMEI}{International Mobile Station Equipment Identity}
\acrodef{IdP}{Identity Provider}
\acrodef{IDS}{Intrusion Detection System}
\acrodef{ISP}{Internet Service Provider}
\acrodef{LEA}{Law Enforcement Agency}
\acrodef{LCPP}{Lightweight Conditional Privacy Preservation}
\acrodef{LLR}{Log Likelihood Ratio}
\acrodef{LTC}{Long Term Certificate}
\acrodef{LTCA}{Long Term \acs{CA}}
\acrodef{MDS}{Misbehavior Detection Scheme}
\acrodef{PHY}{Physical}
\acrodef{H-LTCA}{Home-LTCA}
\acrodef{F-LTCA}{Foreign-LTCA}
\acrodef{LDAP}{Lightweight Directory Access Protocol}
\acrodef{LBS}{Location Based Service}
\acrodef{LSTM}{Long Short-Term Memory}
\acrodef{LTE}{Long Term Evolution}
\acrodef{LuST}{Luxembourg SUMO Traffic}
\acrodef{MA}{Misbehavior Authority}
\acrodef{MCP}{Maneuver Coordination Protocol}
\acrodef{MCM}{Maneuver Coordination Message}
\acrodef{MCS}{Maneuver Coordination Service}
\acrodef{MAC}{Media Access Control}
\acrodef{MCA}{Message \ac{CA}}
\acrodef{MEA}{Misbehavior Evaluation Authority}
\acrodef{MRM}{Minimum Risk Maneuver}
\acrodef{NN}{Neural Network}
\acrodef{NTP}{Network Time Protocol}
\acrodef{OBU}{On-Board Unit}
\acrodef{OEM}{Original Equipment Manufacturer}
\acrodef{OCSP}{Online Certificate Status Protocol}
\acrodef{PCA}{Pseudonym \acs{CA}}
\acrodef{PDP}{Policy Decision Point}
\acrodef{PEP}{Policy Enforcement Point}
\acrodef{PG}{Performance Gain}
\acrodef{PIR}{Private Information Retrieval}
\acrodef{PKC}{Public Key Cryptography}
\acrodef{PKCS}{Public Key Cryptosystem}
\acrodef{PKI}{Public-Key Infrastructure}
\acrodef{PRECIOSA}{Privacy Enabled Capability in Co-operative Systems and Safety Applications}
\acrodef{PFBCE}{Precision-Focused Binary Cross-Entropy}
\acrodef{PRESERVE}{Preparing Secure Vehicle-to-X Communication Systems}
\acrodef{PRIME}{Platoon Restructuring for Incident Mitigation and Exclusion}
\acrodef{P2P}{peer-to-peer}
\acrodef{PR}{Precision-Recall}
\acrodef{PS}{Participatory Sensing}
\acrodef{RA}{Resolution Authority}
\acrodef{RAM}{Random Access Memory}
\acrodef{RBAC}{Role Based Access Control}
\acrodef{RCA}{Root \acs{CA}}
\acrodef{REST}{Representational State Transfer}
\acrodef{RFF}{Radio Frequency Fingerprinting} 
\acrodef{RSU}{Roadside Unit}
\acrodef{RHyTHM}{RHyTHM}
\acrodef{SAML}{Security Assertion Markup Language}
\acrodef{SAS}{Sample Aggregation Service}
\acrodef{SCMS}{Security Credential Management System}
\acrodef{SCORE@F}{Système COopératif Routier Expérimental Français}
\acrodef{SDSI}{Simple Distributed Security Infrastructure}
\acrodef{SRAAC}{Secure Revocable Anonymous Authenticated Inter-Vehicle Communication}
\acrodef{SeVeCom}{Secure Vehicle Communication}
\acrodef{SAE}{Society of Automotive Engineers}
\acrodef{SIT}{Sichere Informationstechnologie}
\acrodef{SLC}{Short-Lived Certificate}
\acrodef{SoA}{Service-oriented-Approach}
\acrodef{SIFS}{Short Inter Frame Space}
\acrodef{SSO}{Single-Sign-On}
\acrodef{SSL}{Secure Sockets Layer}
\acrodef{SOAP}{Simple Object Access Protocol}
\acrodef{STRP}{Spatial Time Reservation Procedure}
\acrodef{SVM}{Support Vector Machine}
\acrodef{TA}{Transition Area}
\acrodef{TACK}{Temporary Anonymous Certified Key}
\acrodef{TFLite}{TensorFlow Lite}
\acrodef{TS}{Task Service}
\acrodef{TLS}{Transport Layer Security}
\acrodef{ToC}{Transition of Control}
\acrodef{TPM}{Trusted Platform Module}
\acrodef{TPR}{True Positive Rate}
\acrodef{TNR}{True Negative Rate}
\acrodef{TTP}{Trusted Third Party}
\acrodef{TVR}{Ticket Validation Repository}
\acrodef{URI}{Uniform Resource Identifier}
\acrodef{UML}{Unified Modeling Language}
\acrodef{VANET}{Vehicular Ad-hoc Network}
\acrodef{V2I}{Vehicle-to-Infrastructure}
\acrodef{V2V}{Vehicle-to-Vehicle}
\acrodef{V2X}{Vehicle-to-Everything}
\acrodef{VC}{Vehicular Communication}
\acrodef{VM}{Virtual Machine}
\acrodef{VSS}{\ac{VC} Security Subsystem}
\acrodef{WAVE}{Wireless Access in Vehicular Environments}
\acrodef{WSDL}{Web Services Discovery Language}
\acrodef{W3C}{World Wide Web Consortium}
\acrodef{V}{Vehicle}
\acrodef{VANET}{Vehicular Ad-hoc Network}
\acrodef{VLR}{Verifier-Local Revocation}
\acrodef{VPKI}{Vehicular Public-Key Infrastructure}
\acrodef{VM}{Virtual Machine}
\acrodef{WS}{Web Service}
\acrodef{WoT}{Web of Trust}
\acrodef{WSACA}{\ac{WAVE} Service Advertisement \ac{CA}}
\acrodef{XML}{Extensible Markup Language}
\acrodef{XACML}{eXtensible Access Control Markup Language}
\acrodef{ZKP}{Zero Knowledge Proof}
\acrodef{3G}{3rd Generation}
\acrodef{ROC}{Receiver Operating Characteristic}
\acrodef{TP}{True Positive}
\acrodef{FP}{False Positive}
\acrodef{TN}{True Negative}
\acrodef{FN}{False Negative}

%% file: Sections/1_introduction.tex
\section{Introduction}
\label{sec:introduction}

\IEEEPARstart{I}{nternet} of Vehicles (\acs{IoV}) transforms modern transportation, with \ac{V2V} and \ac{V2I} communication enabling cooperative driving applications~\cite{papadimitratos2008secure}. Among the most promising representations of this vision is vehicle platooning. It enables multiple vehicles to travel in close formation with coordinated inter-vehicular spacing, yielding substantial improvements in traffic throughput, fuel efficiency, and safety~\cite{amoozadeh2015platoon, amoozadeh2015security, kalogiannis2024prime, li2025attentionguard, abdelmaguid2025beyond}. 

Vehicular platoons operate through \ac{CACC} systems that extend traditional \ac{ACC} capabilities by incorporating real-time communication (\ac{CAM}) of kinematic data, including position, velocity, and acceleration via standardized protocols such as IEEE 802.11p or Cellular \ac{V2X}~\cite{papadimitratos2008secure}. 

This fundamental reliance on continuous wireless information exchange, however, introduces critical security vulnerabilities that threaten operational integrity and passenger safety~\cite{kalogiannis2022attack, li2025attentionguard}. The distributed nature of platoon coordination creates an attack surface where malicious or compromised vehicles can inject falsified information to manipulate the behavior of other platoon members. Such attack consequences range from reduced stability to complete platoon destabilization and increased collision risks~\cite{heijden2017Analyzing}. The effects can be amplified when misbehavior originates from the platoon leader, given its key role in platooning. Unlike external adversaries, which can be thwarted by secure \ac{V2X} protocols~\cite{papadimitratos2008secure, CalandrielloPHL:J:2011, khodaei2018Secmace, KhodaeiP:J:2021b}, internal adversaries transmitting false kinematic data pose a significant challenge that requires real-time detection. 

Traditional approaches to \acp{MDS} in \ac{VANET} have predominantly employed plausibility checks, statistical outlier detection, and rule-based behavioral analysis~\cite{grover2011machine, hakeem2025advancing}. While computationally efficient, these techniques suffer from several limitations: (i) high \ac{FP} rates when encountering sophisticated attacks, (ii) inability to generalize across diverse platoon configurations and control policies, and (iii) inadequate modeling of complex temporal dependencies inherent in multi-vehicle coordination dynamics. The challenge is further compounded by the dynamic nature of platooning scenarios. Topology changes during vehicle \textit{join} and \textit{exit} maneuvers constitute prime moments for attack as they involve multiple cars, potentially in different lanes (e.g., during a join). Such attacks can be catastrophic not only to the platoon itself but also to other road users~\cite{kalogiannis2022attack, kalogiannis2023vulnerability}, posing a two-fold challenge for misbehavior detection during maneuvering: \acp{FP} disrupt legitimate platoon operations, while \acp{FN} enable successful attacks.

Several detection approaches emerged to address these limitations through diverse methodologies. Probabilistic methods employing \ac{GMM}-\ac{HMM} achieve 87\% F1-score~\cite{kalogiannis2022attack}, specialized classifiers leverage \ac{RF} for rapid inference~\cite{boddupalli2021replace, boddupalli2019redem, mousavinejad2022secure}, and recurrent networks (\ac{LSTM}/\ac{GRU}) demonstrate improved accuracy~\cite{wang2021deep, ko2021approach}. Most closely related to our work, AttentionGuard~\cite{li2025attentionguard} applies attention mechanisms to achieve 88-92\% accuracy with 96-99\% \ac{AUC}. However, three essential constraints persist across these approaches: (i) inference latency incompatible with real-time requirements; (ii) reliance on manual feature engineering or position-specific architectures limiting cross-configuration generalization; (iii) sequential processing in recurrent networks. The latter prevents efficient parallel training while struggling to simultaneously capture both intra-vehicle temporal evolution and inter-vehicle spatial correlations, which are essential for detecting coordinated attacks during topology changes.

\rev{Transformer architectures~\cite{vaswani2017attention} can address constraints (ii) and (iii) through self-attention mechanisms that enable parallel sequence processing while modeling intra-vehicle temporal dynamics. By reshaping the input to attend across both vehicle and time dimensions, a spatio-temporal variant further captures inter-vehicle spatial correlations. The former enables efficient vehicle-local deployment, while the latter captures attack propagation patterns across platoon formations. Both prove essential for distinguishing legitimate coordination dynamics from malicious manipulation, thereby minimizing \acp{FP} during legitimate maneuvers while maintaining robust detection of sophisticated, coordinated attacks.}

Concurrently, the maturation of edge computing within vehicular networks enables deploying \ac{ML} models closer to data sources, addressing the ultra-low latency requirements (typically under 100$ms$) for safety-critical driving applications~\cite{shih2018architectural}. Edge \ac{AI} enables real-time inference on resource-constrained platforms while preserving privacy through localized processing and reducing communication overhead to cloud infrastructure~\cite{zhang2020mobile}.

\rev{To address these challenges, we present \textsc{AIMformer}, a transformer-based framework for real-time misbehavior detection in vehicular platoons with edge deployment capabilities. Our approach supports two operational modes: a \emph{$B \cdot V$} configuration where self-attention captures intra-vehicle temporal dynamics independently per vehicle (enabling vehicle-local deployment), and a \emph{$V \cdot T$} configuration where attention operates across the concatenated vehicle-time dimension to model inter-vehicle spatial correlations (suited for infrastructure-based deployment). We summarize our \textbf{contributions} as follows:}

\begin{enumerate}[leftmargin=*]
    \item We build on our prior work~\cite{li2025attentionguard} and develop the first transformer-based misbehavior detection framework specifically designed for vehicular platooning environments, incorporating global positional encoding mechanisms that maintain temporal awareness across multi-vehicle sequences with asynchronous dynamics.
    \item We propose a novel \ac{PFBCE} loss function that addresses the precision-recall trade-off inherent in security-critical applications by introducing parameterized penalties for \ac{FP} predictions while preserving attack detection capability through balanced positive class weighting.
    \item We conduct an extensive comparative analysis across 4 different platoon controllers, various attack types (i.e., position, speed, acceleration falsification with constant, and gradual offsets), and mobility scenarios (e.g., join, exit, steady-state), demonstrating superior performance ($\geq$ 0.93) against baseline architectures including \ac{LSTM}, BiLSTM, \ac{CNN}, \ac{CNN}-\ac{LSTM}, \ac{GRU}, and \ac{MLP}.
    \item We demonstrate the practical viability of edge deployment through comprehensive evaluation and analysis using popular frameworks (e.g., \ac{TFLite}, \ac{ONNX}, TensorRT), achieving sub-millisecond inference latencies suitable for real-time deployment on resource-constrained embedded platforms.
\end{enumerate}

\acused{IoV}

\textbf{Paper Organization.} Sec.~\ref{sec:preliminaries} establishes preliminaries including \ac{IoV} platooning architectures, edge \ac{AI}, and transformer encoder. Sec.~\ref{sec:system_and_adversary} formalizes our system and adversarial models. Sec.~\ref{sec:implemenation} details the \textsc{AIMformer} architecture, feature engineering pipeline, training methodology, and deployment optimization strategies. Sec.~\ref{sec:performance} presents performance evaluation including ablation studies, baseline comparisons, and deployment analysis on edge hardware. \rev{Sec.~\ref{sec:discussion} discusses threshold selection, deployment strategies, generalization, and limitations.} Sec.~\ref{sec:related_work} surveys related work in vehicular misbehavior detection, transformer applications in security domains, and edge \ac{AI} for \ac{ITS}. Sec.~\ref{sec:conclusion} concludes with future research directions.

%% file: Sections/2_background.tex
\section{Preliminaries}
\label{sec:preliminaries}

This section establishes the foundational concepts necessary for understanding the proposed framework: \ac{IoV} platooning coordination, transformer encoder architecture, and the context of Edge \ac{AI} deployment.

\subsection{Vehicular Platooning and V2V/V2I Communication}

Vehicle platooning enables coordinated convoy formations through \ac{V2V} and \ac{V2I} communication, where \ac{CACC} systems extend traditional \ac{ACC} by exchanging real-time kinematic data (i.e., position, velocity, acceleration) via IEEE 802.11p or cellular \ac{V2X} protocols.

\textit{Communication Topologies and Control Strategies:} Platoon communication architectures operate through distinct topological configurations that determine information flow patterns. The predominant topologies include: \textit{Predecessor-Following}, where vehicles receive information exclusively from their immediate predecessor, minimizing communication overhead but potentially suffering from error propagation; \textit{Predecessor-Leader}, where vehicles maintain links with both predecessor and platoon leader, enhancing string stability through dual information sources; and \textit{Bidirectional Communication}, where vehicles exchange information with both preceding and following vehicles, facilitating enhanced situational awareness and disturbance rejection capabilities.

Platoon coordination employs two primary spacing policies: \textit{\ac{CTH}} maintains time-based following distances that scale with vehicle speed, enhancing stability at higher velocities while reducing road utilization efficiency; \textit{\ac{CVS}} enforces fixed physical distances regardless of speed, maximizing road capacity but requiring precise control mechanisms.

Several control strategies implement these policies: PATH (CVS with Predecessor-Leader topology), Ploeg (CTH with Predecessor-Following topology), Consensus (hybrid policy with multi-source information), and Flatbed (CVS with leader and predecessor speed information). These different information flows and controller policies lead to varied vulnerability susceptibility and platoon reactions~\cite{heijden2017Analyzing, kalogiannis2022attack, hendrix2024platoon}.

\subsection{Transformer Encoder Architecture}

The Transformer encoder utilizes self-attention mechanisms for parallel sequence processing, capturing long-range dependencies without the computational limitations inherent in \acp{RNN}.

\textbf{Self-Attention Mechanism.} The self-attention mechanism operates by transforming input sequences into three fundamental representations: queries ($Q$), keys ($K$), and values ($V$). For an input sequence $x_1, x_2, \ldots, x_n$ where each $x_i \in \mathbb{R}^d$, the scaled dot-product attention computes:

{\footnotesize
\begin{equation}
    \text{Attention}(Q,K,V) = \text{softmax}\left(\frac{QK^T}{\sqrt{d}}\right)V
\end{equation}
}

This formulation enables each sequence position to attend to all other positions simultaneously, creating a comprehensive representational model that captures complex temporal dependencies essential for time-series analysis tasks.

\textbf{Multi-Head Attention and Positional Encoding.} Multi-head attention extends the self-attention mechanism by computing multiple attention operations in parallel across different representational subspaces. This parallel processing enables the model to capture diverse types of dependencies simultaneously, with different attention heads specializing in distinct relational patterns within the data. The multi-head attention mechanism is defined as:

{\footnotesize
\begin{equation}
\text{MultiHead}(Q,K,V) = \text{Concat}(\text{head}_1, \ldots, \text{head}_h) W^O
\end{equation}
}

where each attention head operates in a lower-dimensional subspace, allowing the model to focus on different aspects of the input sequence concurrently. Since the attention mechanism lacks inherent sequential ordering information, positional encoding provides the model with explicit positional awareness. The Transformer employs sinusoidal positional encodings that are added directly to input embeddings, enabling the model to distinguish between temporal positions while generalizing to sequences of varying lengths.

\rev{In our platoon setting, $Q$, $K$, and $V$ are learned projections of vehicle kinematic features at each timestep. In the per-vehicle ($B \cdot V$) configuration, self-attention operates within a single vehicle's time window to capture temporal patterns; in the spatio-temporal ($V \cdot T$) configuration, it spans the joint vehicle-time dimension, enabling cross-vehicle attention for improved attack detection and attacker localization.}

\subsection{Edge AI Deployment Context}

Edge \ac{AI} enables \ac{ML} model execution on devices proximate to data sources, addressing latency, bandwidth, and privacy constraints inherent in cloud-centric approaches~\cite{filho2022systematic, murshed2021machine}. For safety-critical vehicular applications requiring sub-100$ms$ response times, edge deployment facilitates real-time misbehavior detection while preserving the benefits of localized processing. Recent embedded implementations demonstrate \ac{DNN} architectures that achieve high accuracy on resource-constrained hardware~\cite{2022_jamming_hussain, hussain2024edge, 2025_lightweight_edge_ai_rff}, making distributed security frameworks feasible for vehicular platoons. Within that context, edge \ac{AI} enables distributed misbehavior detection, real-time trust management, and adaptive coordination that leverage local processing while minimizing communication overhead to centralized infrastructure.

%% file: Sections/3_adversary_system.tex
\section{System and Adversarial Model}
\label{sec:system_and_adversary}

\begin{figure}[!t]
  \centering
    \begin{subfigure}[b]{\linewidth}
        \centering
        \includegraphics[width=\linewidth]{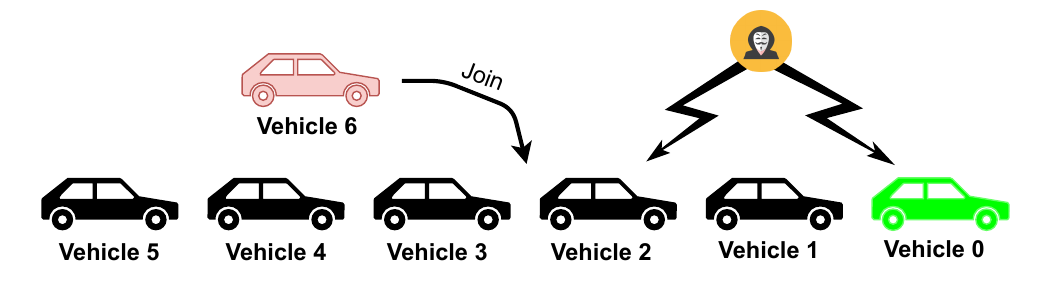}
        \caption{}
        \label{fig:adv_a}
    \end{subfigure}
    \vfill
    \begin{subfigure}[b]{\linewidth}
    \centering
    \includegraphics[width=\linewidth]{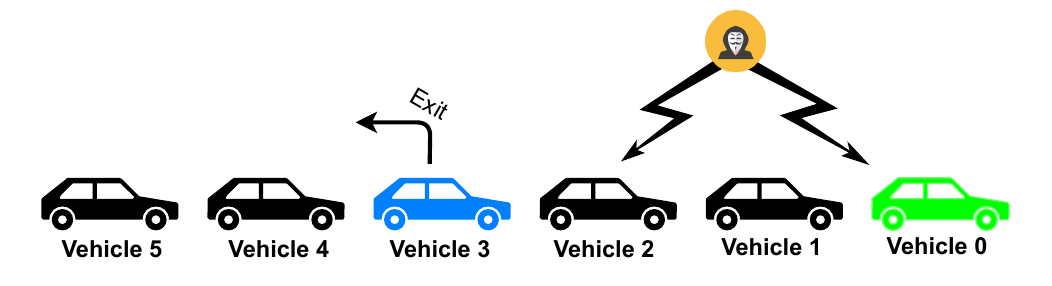}
    \caption{}
    \label{fig:adv_b}
    \end{subfigure}
    
    \vfill
    \begin{subfigure}[b]{\linewidth}
    \centering
    \includegraphics[width=\linewidth]{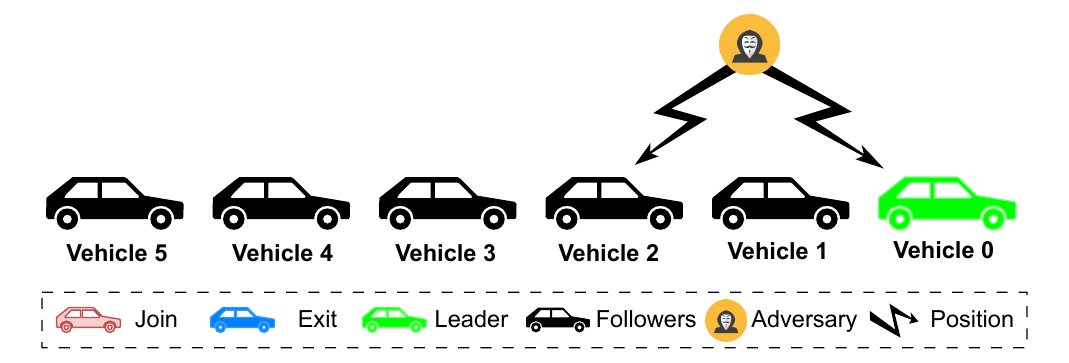}
    \caption{}
    \label{fig:adv_c}
    \end{subfigure}
    \caption{Attacks during platoon topology changes. \textbf{(a)} \emph{Join} maneuver: a new vehicle (Vehicle 6) attempts to merge into the platoon while an adversary forges/perturbs kinematic information. \textbf{(b)} \emph{Exit} maneuver: a member (Vehicle 3) leaves the platoon while the adversary initiates the message manipulation. \textbf{(c)} \emph{Steady-State}: the attacker performs a falsification attack on the platoon during normal cruising.}
    \label{fig:adversarial-models}
\end{figure}

\subsection{System Model}
\label{subsec:e2e_attack_v2x_detection_pipeline}

We consider vehicles implementing \ac{V2X} protocol stacks~\cite{PapadimitratosFEBC:J:2009} equipped with valid \ac{VC} cryptographic credentials obtained through legitimate vehicle registration, allowing them to participate in \ac{V2X} communications and platooning operations~\cite{1609, khodaei2018Secmace, KhodaeiP:J:2021b, KhodaeiNP:J:2023}. Successful misbehavior detection can trigger the credential revocation mechanisms, preventing future malicious participation in platoon operations. Platoons operate under a designated leader who coordinates vehicle enrollment and departure occurring through individual vehicle join/exit maneuvers or platoon merging operations. Upon admission, joining vehicles position themselves at leader-designated locations, and the updated configuration is broadcast to all members. Our system model assumes a string-stable platoon of vehicles traveling on a highway segment, allowing for join/exit maneuvers at any position based on vehicle capabilities and destination compatibility.

\begin{figure*}[!htbp]
    \centering
    \includegraphics[width=0.93\textwidth]{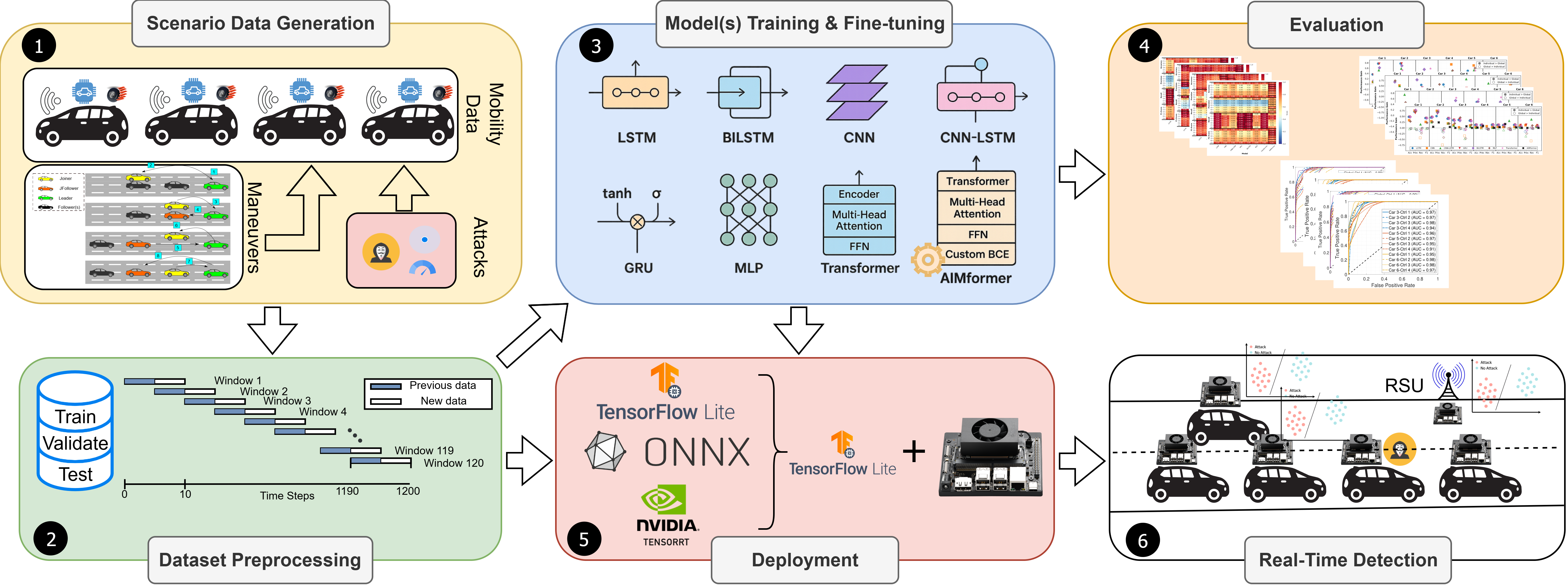}
    \caption{Pipeline for edge AI-based V2X attack detection: \protect\blackcircled{\textbf{1}} Scenario data generation with mobility traces, maneuvers, and attacks; \protect\blackcircled{\textbf{2}} dataset preprocessing and temporal windowing into train/validation/test splits; \protect\blackcircled{\textbf{3}} model training with \acs{LSTM}, \acs{BiLSTM}, \acs{CNN}, \acs{CNN}-\acs{LSTM}, \acs{GRU}, \acs{MLP}, and Transformer/\textsc{AIMformer} (custom \ac{BCE}); \protect\blackcircled{\textbf{4}} offline evaluation (confusion matrices, metrics, ROC); \protect\blackcircled{\textbf{5}} deployment via \ac{TFLite}, \ac{ONNX}, and TensorRT to embedded hardware; and \protect\blackcircled{\textbf{6}} real-time detection on vehicles/RSUs. Arrows indicate the end-to-end workflow.}
    \label{fig:pipeline}
\end{figure*}

\subsection{Adversarial Models}
\label{subsec:adversary-model}

We consider internal adversaries equipped with valid cryptographic credentials, obtained through legitimate vehicle registration. Following the general adversary model for secure and privacy-preserving \ac{VC} systems~\cite{papadimitratos2008secure, PapadimitratosGH:C:2006} and vehicular platooning assumptions~\cite{amoozadeh2015security, heijden2017Analyzing, dantas2020formal, ucar2018ieee}, these authenticated attackers can participate in \ac{V2X} communications and transmit messages accepted as legitimate by other platoon members. However,
internal adversaries cannot extract private keys that are stored in \acp{HSM}~\cite{papadimitratos2008secure}, provisioned during the vehicle registration~\cite{khodaei2018Secmace}. Adversarial behavior stems from direct vehicle control, or malware intercepting internal data streams (e.g., \ac{CAN} bus messages, sensor outputs) before \ac{V2X} transmission.

Internal adversaries mount falsification attacks by manipulating kinematic data in transmitted \acp{CAM}. 
Specifically, we consider three data falsification attacks executed by rational adversaries who maximize attack effectiveness while minimizing self-harm~\cite{kalogiannis2022attack}. First, attackers can broadcast incorrect positional coordinates, misleading neighboring vehicles about their actual location. This attack type particularly affects controllers that rely on absolute positioning information for distance calculations and coordination decisions. Second, malicious vehicles can report incorrect velocity values, causing platoon followers to make inappropriate acceleration decisions. The severity of impact varies based on control strategy, with \ac{CVS}-based controllers showing greater vulnerability to speed-based attacks. Finally, attackers can transmit false acceleration data. This attack type can cause immediate and severe destabilization effects across diverse controller implementations.

\textbf{Attack Windows.} The topology reconfiguration events represent critical vulnerability windows: during join and exit maneuvers, platoons experience temporary instability characterized by modified communication topologies, redistributed control responsibilities, and heightened requirements for accurate state information exchange. Adversaries exploit these events to amplify attack effectiveness and maximize disruption.

\textbf{Attack Scenarios.} Fig.~\ref{fig:adversarial-models} illustrates three attack scenarios: \textit{(a) Join Maneuver} where Vehicle $6$ integrates while the adversary broadcasts falsified kinematic data, targeting the admission and spacing phase to create unsafe gaps or collision risks; \textit{(b) Exit Maneuver} where the adversary manipulates \acp{CAM} during Vehicle $3$'s departure, exploiting the critical period when remaining vehicles reestablish predecessor-follower relationships; and \textit{(c) Steady-State} where the adversary performs data falsification during normal cruising operations without topology changes. For each attack scenario, we consider both an attacker leading the platoon and an attacker traveling just ahead of the maneuvering position. This setup ensures both maximal attack reach (leader attacker) and amplified attack impact (vehicle 2 attacker) by potentially causing collisions in both lanes (as the maneuvering vehicle changes lanes)~\cite{kalogiannis2022attack}.

We examine $9$ distinct attack types spanning three manipulation strategies~\cite{kalogiannis2022attack}: \textit{Constant Offset Attacks} introducing fixed biases to position, velocity, or acceleration; \textit{Gradual Offset Attacks} applying time-varying drift to kinematic parameters; and \textit{Combined Physics-Consistent Attacks} simultaneously falsifying multiple state variables while maintaining kinematic feasibility to evade simple plausibility checks. \rev{Table~\ref{tab:attack_params} summarizes the parameter ranges for all attack types and simulation assumptions. For combined attacks, the drift of each value also drifts the others in a kinematically consistent way.}

\rev{
\begin{table}[!htbp]
\centering
\caption{\rev{Attack, simulation, and preprocessing parameters.}}
\label{tab:attack_params}
\footnotesize
\resizebox{\columnwidth}{!}{%
\begin{tabular}{|l|l|}
\hline
\textbf{Parameter} & \textbf{Value / Range} \\
\hline
\hline
\multicolumn{2}{|c|}{\textbf{Attack Parameters}~\cite{kalogiannis2022attack}} \\
\hline
Constant position offset & $\{3, 5, 7, 9, 11\}$ m \\
Constant speed offset & $\{-50, 0, 50, 100, 150\}$ km/h \\
Constant acceleration offset & $\{-30, -10, 0, 10, 30\}$ m/s$^{2}$ \\
Gradual position drift & $[-10, 40]$ m \\
Gradual speed drift & $[-10, 17]$ m/s \\
Gradual acceleration drift & $[-10, 10]$ m/s$^{2}$ \\
Combined attack & $[-10, 10]$ in pos (m), speed (m/s), accel (m/s$^{2}$) \\ 
Sensor error (V2V) & $\epsilon_p{=}1$ m, $\epsilon_s{=}0.1$ m/s, $\epsilon_a{=}0.01$ m/s$^{2}$ \\
Sensor error (Radar) & $\epsilon_p{=}0.1$ m, $\epsilon_s{=}0.1$ m/s \\
\hline
\hline
\multicolumn{2}{|c|}{\textbf{Simulation Settings}} \\
\hline
Simulator & SUMO + Plexe \\
Platoon size & 6 vehicles \\
Platoon speeds & 50, 80, 100, 150 kmph\\
Controllers (spacing) & PATH (5m), Ploeg (0.5s), Consensus (0.8s), Flatbed (5m) \\
CAM frequency & 10 Hz (100$ms$ interval) \\
Communication & IEEE 802.11p\\
Maneuvers & Join, Exit, Steady-state \\
Attacker positions & Leader (pos. 0), Follower (pos. 2) \\
\hline
\hline
\multicolumn{2}{|c|}{\textbf{Preprocessing}} \\
\hline
Window size ($T$) & 10 timesteps (1 second) \\
Stride & 1 timestep (100$ms$) \\
Features ($F$) & 7 (pos, speed, accel, $\times$ self/pred/leader) \\
Normalization & Global z-score per feature \\
\hline
\end{tabular}%
}
\end{table}
}

%% file: Sections/4_implementaion.tex
\section{Proposed Framework and Implementation}
\label{sec:implemenation}

\subsection{Proposed Framework}
Fig.~\ref{fig:pipeline} presents our end-to-end attack detection pipeline from data generation to deployment. In step \blackcircled{\textbf{1}}, we generate scenario data incorporating diverse mobility patterns (middle-join, middle-exit, normal traveling) across different platoon formations with attackers in various positions (described in Sec.~\ref{subsec:adversary-model}). The mobility features are sourced from ego-vehicle sensors, each with its own sensor errors, capturing the resulting mobility effects after controller actuation. The deployed \ac{MDS} learns to distinguish misbehavior from nominal mobility based on these patterns. Step \blackcircled{\textbf{2}} preprocesses the mobility data into normalized train/validation/test sets, segmenting them into windows of ten messages (1 second) with 100$ms$ step size, enabling inference after each disseminated \ac{CAM}. Each timestep receives a label (attack/no attack) and mask (Eq.~\ref{eq:m_pad}).

Step \blackcircled{\textbf{3}} involves training baseline architectures (\ac{LSTM}, \ac{BiLSTM}, \ac{CNN}, \ac{CNN}-\ac{LSTM}, \ac{GRU}, \ac{MLP}, Transformer) and our \textsc{AIMformer} solution, creating both global models (deployable in \acp{RSU}) and individual vehicle-specific models (Sec.~\ref{subsec:model_training}). Step \blackcircled{\textbf{4}} evaluates all models on the unseen test set, applying iterative refinement (Sec.~\ref{sec:performance}). Step \blackcircled{\textbf{5}} converts and quantizes models via \ac{TFLite}, \ac{ONNX}, and TensorRT for resource-constrained deployment, evaluating inference time and memory footprint (Sec.~\ref{subsec:deployment-evaluation}). Finally, step \blackcircled{\textbf{6}} analyzes individual versus global model performance across platoon positions (Sec.~\ref{subsec:model_comparison}). 

\rev{
\begin{table}[!htbp]
\centering
\caption{\rev{Input tensor dimensions and offset convention.}}
\label{tab:tensor-dims}
\footnotesize
\resizebox{\columnwidth}{!}{
\begin{tabular}{|c|l|l|}
\hline
\textbf{Symbol} & \textbf{Description} & \textbf{Typical Value} \\
\hline
$B$ & Batch size & 128 \\
$V$ & Number of vehicles in the platoon & 6 (global), 1 (individual) \\
$T$ & Sequence length (timesteps per window) & 10 (= 1 second) \\
$F$ & Feature dimension per timestep & 7 \\
$\text{offset}_v$ & Global temporal offset for vehicle $v$, & Integer $\geq 0$ \\
\hline
$d_{model}$ & Hidden dimension of the model & 128 \\
$h$ & Number of attention heads & 2 \\
$d_k$ & Key/query dimension per head ($d_{model}/h$) & 64 \\
$d_{ff}$ & Feed-forward network dimension ($4 \cdot d_{model}$) & 512 \\
$L$ & Number of encoder blocks & 4 (global), 2 (individual) \\
\hline
\end{tabular}
}
\end{table}
}

\begin{table}[!htbp]
\centering
\caption{Notation for the Transformer-encoder.}
\label{tab:notation-architecture}
\resizebox{0.9\columnwidth}{!}{%
    \begin{tabular}{|p{3.5cm}|l|}
    \hline
    \textbf{Symbol} & \textbf{Description}\\
    \hline
    \hline
    \multicolumn{2}{|c|}{\textbf{Model Parameters}} \\
    \hline
    $\mathbf{W}_{embed}$ & Input embedding weight matrix \\
    $\mathbf{b}_{embed}$ & Input embedding bias vector  \\
    $\mathbf{W}_i^Q$ & Query projection matrix for head $i$ \\
    $\mathbf{W}_i^K$ & Key projection matrix for head $i$  \\
    $\mathbf{W}_i^V$ & Value projection matrix for head $i$  \\
    $\mathbf{W}^O$ & Multi-head output projection matrix \\
    $\mathbf{W}_1$ & First feed-forward weight matrix \\
    $\mathbf{W}_2$ & Second feed-forward weight matrix  \\
    $\mathbf{b}_1$ & First feed-forward bias vector  \\
    $\mathbf{b}_2$ & Second feed-forward bias vector \\
    $\mathbf{W}_{out}$ & Output projection weight matrix \\
    $\mathbf{b}_{out}$ & Output projection bias \\
    \hline
    \hline
    \multicolumn{2}{|c|}{\textbf{Hidden States and Intermediate Tensors}} \\
    \hline
    $\mathbf{H}^{(0)}$ & Initial embedded representations \\
    $\tilde{\mathbf{H}}^{(\ell)}$ & Hidden states after layer $\ell$ \\
    $\mathbf{A}^{(\ell)}$ & Attention output at layer $\ell$ \\
    $\mathbf{H}_1^{(\ell)}$ & Intermediate hidden states at layer $\ell$  \\
    $\mathbf{F}^{(\ell)}$ & Feed-forward output at layer $\ell$ \\
    $\mathbf{O}_{flat}$ & Flattened output before reshaping \\
    \hline
    \hline
    \multicolumn{2}{|c|}{\textbf{Attention Mechanism}} \\
    \hline
    $\mathbf{Q}$ & Query matrix \\
    $\mathbf{K}$ & Key matrix \\
    $\mathbf{V}$ & Value matrix \\
    $\text{head}_i$ & Output of attention head $i$ \\
    $\rev{\mathbf{M} = \mathbf{M}_{pad}}$ & \rev{Attention mask = Padding mask}  \\
    \hline
    \end{tabular}%
}
\end{table}

\subsection{Model Design}
\label{se:Model_Design}

\textbf{Model Architecture.} We present a Transformer-based architecture for single- or multi-vehicle binary classification. Given input $\mathbf{X} \in \mathbb{R}^{B \times V \times T \times F}$ (Table~\ref{tab:tensor-dims}), the model produces binary predictions $\hat{\mathbf{Y}} \in \mathbb{R}^{B \times V \times T}$ for each timestep. Global positional encoding with vehicle-specific temporal offsets aligns the vehicles individual time windows to the vehicle trip:

{\footnotesize
\begin{align}
\mathbf{P}_{v,t} &= PE(g_{v,t}) \label{eq:capital_position} \\
g_{v,t} &= t + \text{offset}_v \label{eq:global_offset}
\end{align}
}

\noindent where $g_{v,t}$ represents the global time position for vehicle $v$ at local timestep $t$. Input processing reshapes and embeds multi-vehicle data for parallel computation:

{\footnotesize
\begin{align}
\label{eq:reshape}
\mathbf{X}_{flat} &= \text{reshape}(\mathbf{X}, [B \cdot V, T, F]) \\
\mathbf{H}^{(0)} &= \mathbf{X}_{flat} \mathbf{W}_{embed} + \mathbf{b}_{embed} \\
\tilde{\mathbf{H}}^{(0)} &= \mathbf{H}^{(0)} + \mathbf{P}
\end{align}
}

\noindent where $\mathbf{W}_{embed} \in \mathbb{R}^{F \times d_{model}}$, $\mathbf{b}_{embed} \in \mathbb{R}^{d_{model}}$ are embedding parameters, and $\mathbf{P} \in \mathbb{R}^{B \cdot V \times T \times d_{model}}$ contains positional encodings. Multi-head self-attention is formulated as:

{\footnotesize
\begin{align}
\text{MultiHead}(\mathbf{Q}, \mathbf{K}, \mathbf{V}) &= \text{Concat}(\text{head}_1, \ldots, \text{head}_h) \mathbf{W}^O
\end{align}
\begin{align}
\text{head}_i &= \text{Attention}(\mathbf{Q}\mathbf{W}_i^Q, \mathbf{K}\mathbf{W}_i^K, \mathbf{V}\mathbf{W}_i^V)
\end{align}
\begin{align}
\text{Attention}(\mathbf{Q}, \mathbf{K}, \mathbf{V}) &= \text{softmax}\left(\frac{\mathbf{Q}\mathbf{K}^T}{\sqrt{d_k}} + \mathbf{M}\right) \mathbf{V}
\end{align}
}
where $\mathbf{W}_i^Q, \mathbf{W}_i^K, \mathbf{W}_i^V \in \mathbb{R}^{d_{model} \times d_k}$ are projection matrices, $d_k = d_{model}/h$, and $\mathbf{M}$ is the attention mask (Table~\ref{tab:notation-architecture}). \rev{The masking method handles variable-length sequences with asynchronous entry/exit times through a padding mask:}

{\footnotesize
\begin{align}
\label{eq:m_pad}
\mathbf{M} &= \mathbf{M}_{pad}[i,j] = \begin{cases}
0, & \text{if position } j \text{ is valid for sequence } i \\
-\infty, & \text{if position } j \text{ is padded for sequence } i
\end{cases}
\end{align}
}
\rev{The padding mask $\mathbf{M}_{pad}$ prevents attention to invalid (padded) positions arising from vehicles with different observation periods due to join/exit events. Unlike autoregressive language models, our detection task benefits from bidirectional attention across the full observation window; causal masking is therefore intentionally omitted.} 
Each encoder block $\ell \in \{1, 2, \ldots, L\}$ applies standard Transformer operations with residual connections and layer normalization:

{\footnotesize
\begin{align}
\mathbf{A}^{(\ell)} &= \text{MultiHead}(\tilde{\mathbf{H}}^{(\ell-1)}, \tilde{\mathbf{H}}^{(\ell-1)}, \tilde{\mathbf{H}}^{(\ell-1)}) \\
\mathbf{H}_1^{(\ell)} &= \text{LayerNorm}(\tilde{\mathbf{H}}^{(\ell-1)} + \text{Dropout}(\mathbf{A}^{(\ell)})) \\
\mathbf{F}^{(\ell)} &= \text{FFN}(\mathbf{H}_1^{(\ell)}) \\
\tilde{\mathbf{H}}^{(\ell)} &= \text{LayerNorm}(\mathbf{H}_1^{(\ell)} + \text{Dropout}(\mathbf{F}^{(\ell)}))
\end{align}
}

where the feed-forward network is:

{\footnotesize
\begin{align}
\text{FFN}(\mathbf{x}) &= \text{ReLU}(\mathbf{x}\mathbf{W}_1 + \mathbf{b}_1)\mathbf{W}_2 + \mathbf{b}_2
\end{align}
}
with $\mathbf{W}_1 \in \mathbb{R}^{d_{model} \times d_{ff}}$, $\mathbf{W}_2 \in \mathbb{R}^{d_{ff} \times d_{model}}$, and typically $d_{ff} = 4 \cdot d_{model}$. Final output projection recovers the multi-vehicle structure:

{\footnotesize
\begin{align}
\mathbf{O}_{flat} &= \tilde{\mathbf{H}}^{(L)} \mathbf{W}_{out} + \mathbf{b}_{out}
\end{align}
}

{\footnotesize
\begin{align}
    \hat{\mathbf{Y}} &= \text{reshape}(\mathbf{O}_{flat}, [B, V, T])
\end{align}
}
where $\mathbf{W}_{out} \in \mathbb{R}^{d_{model} \times 1}$ and $\mathbf{b}_{out} \in \mathbb{R}$ are binary classification parameters. \rev{The output $\hat{\mathbf{Y}}$ contains raw logits; a sigmoid activation $\sigma(\cdot)$ is applied to obtain per-timestep attack probabilities, and a timestep is classified as an attack if $\sigma(\hat{y}_{i,j}) > \tau$.} 
\rev{This default architecture, referred to as the \emph{$B \cdot V$ configuration}, reshapes the input to $[B \cdot V, T, F]$ (Eq.~\ref{eq:reshape}), so self-attention operates independently over each vehicle's temporal sequence. $B \cdot V$ does \emph{not} model inter-vehicle spatial correlations; it captures intra-vehicle temporal dynamics, with global positional encoding (Eqs.~\ref{eq:capital_position}--\ref{eq:global_offset}) providing trajectory-phase awareness. This vehicle-independent processing enables flexible deployment as either global (multi-vehicle) or individual (single-vehicle) models without architectural modifications. This allows local vehicle execution since it requires no observations from other platoon members.}

\rev{An alternative \emph{$V \cdot T$ configuration} reshapes the input to $[B, V \cdot T, F]$, enabling self-attention across the concatenated vehicle-time dimension to capture inter-vehicle spatial correlations alongside temporal dynamics. This variant requires whole-platoon observations at both training and inference, making it suitable for infrastructure-based (e.g., \ac{RSU}) deployment. Sec.~\ref{subsec:inter-vehicle-aware-endoding} evaluates $V \cdot T$ and compares it with $B \cdot V$.}

\begin{table}[!htbp]
\centering
\caption{Mathematical notation for \ac{PFBCE} Loss function.}
\label{tab:notation-loss}
\resizebox{0.85\columnwidth}{!}{%
    \begin{tabular}{|c|l|}
    \hline
    \textbf{Symbol} & \textbf{Description} \\
    \hline
    \multicolumn{2}{|c|}{\textbf{Input Variables}} \\
    \hline
    $y_{i,j}$ & Ground truth binary label for vehicle i at time step j\\
    $\hat{y}_{i,j}$ & Predicted logit for vehicle i at time step j\\
    $m_{i,j}$ & Binary loss mask for vehicle i at time step j\\
    \hline
    \multicolumn{2}{|c|}{\textbf{Sets and Derived Variables}} \\
    \hline
    $\mathcal{M}$ & Set of valid vehicle-time observations, $\{(i,j) : m_{i,j} = 1\}$\\
    $|\mathcal{M}|$ & Number of valid vehicle-time observations\\
    $\ell_{i,j}$ & Standard \ac{BCE} loss for vehicle $i$ at time step $j$\\
    $w_{\text{FP}}^{(i,j)}$ & \ac{FP} penalty weight for vehicle $i$ at time step $j$\\
    $w_{\text{pos}}^{(i,j)}$ & Positive class weight for vehicle $i$ at time step $j$\\
    \hline
    \multicolumn{2}{|c|}{\textbf{Hyperparameters}} \\
    \hline
    $\lambda_{\text{FP}}$ & \ac{FP} penalty weight \\
    $\lambda_{\text{pos}}$ & Positive class weight \\
    $\tau$ & \ac{FP} threshold\\
    $\epsilon$ & Numerical stability constant\\
    \hline
    \multicolumn{2}{|c|}{\textbf{Functions}} \\
    \hline
    $\sigma(\cdot)$ & Sigmoid activation function, $\sigma(z) = \frac{1}{1 + e^{-z}}$ \\
    $\mathcal{L}_{\text{PFBCE}}(\cdot)$ & Precision-focused \ac{BCE} loss \\
    \hline
    \end{tabular}%
}
\end{table}
\textbf{Loss Function.} We propose \ac{PFBCE} to enhance precision by explicitly penalizing \acp{FP} while maintaining balanced treatment of positive samples, enabling detection of attack-influenced behavior that mimics nominal mobility~\cite{li2025attentionguard}:

{\footnotesize
\begin{align}
\label{eq:loss}
\mathcal{L}_{\text{PFBCE}}(\mathbf{y}, \hat{\mathbf{y}}) &= \frac{1}{|\mathcal{M}|} \sum_{i,j \in \mathcal{M}} \ell_{i,j} \cdot w_{\text{FP}}^{(i,j)} \cdot w_{\text{pos}}^{(i,j)} \cdot m_{i,j}
\end{align}
}

{\footnotesize
\begin{align}
\ell_{i,j} &= -y_{i,j} \log(\sigma(\hat{y}_{i,j})) - (1 - y_{i,j}) \log(1 - \sigma(\hat{y}_{i,j})) \\
w_{\text{FP}}^{(i,j)} &= \begin{cases} 
\lambda_{\text{FP}}, & \text{if } y_{i,j} = 0 \text{ and } \sigma(\hat{y}_{i,j}) > \tau \\
1, & \text{otherwise}
\end{cases} \\
\label{eq:pos_weight}
w_{\text{pos}}^{(i,j)} &= \begin{cases} 
\lambda_{\text{pos}}, & \text{if } y_{i,j} = 1 \\
1, & \text{if } y_{i,j} = 0
\end{cases} \\
|\mathcal{M}| &= \sum_{i,j} m_{i,j} + \epsilon
\end{align}
}

The \ac{FP} penalty $w_{\text{FP}}^{(i,j)}$ penalizes positive predictions ($\sigma(\hat{y}_{i,j})>\tau$) on benign samples, encouraging conservative classification. The positive class weight $w_{\text{pos}}^{(i,j)}$ addresses class imbalance through moderate upweighting ($\lambda_{\text{pos}}$), maintaining sensitivity while benefiting from precision enhancement. The mask $m_{i,j}$ excludes invalid data (different entry/exit times or inapplicable scenarios), with normalization by $|\mathcal{M}|$ ensuring consistent loss magnitude (Table~\ref{tab:notation-loss}). 

\textbf{Why PFBCE Over Focal Loss.} Focal Loss emphasizes hard examples through $(1 - p_t)^\gamma$, but treats all difficult samples uniformly regardless of error type, preventing direct control of the precision-recall trade-off. \ac{PFBCE} instead provides explicit and interpretable error asymmetry: the \ac{FP} penalty $w_{\text{FP}}$ suppresses confident incorrect positives to improve precision, while the positive-class weight $w_{\text{pos}}$ handles class imbalance to tune recall. This decoupled design enables precise control over precision-recall behavior that Focal Loss cannot provide.

\subsection{Training and Tuning}
\label{subsec:model_training}

\textbf{Model Training.} We design two modeling approaches: \textit{global} models trained on all platoon vehicles' data to learn whole-platoon mobility patterns, and \textit{individual} models trained per-vehicle for position-specific local classification.

\textbf{Model Tuning.} We performed hyperparameter tuning using Keras Tuner's Hyperband algorithm~\cite{li2018hyperband}, which balances exploration and computational cost by early termination of poor trials. Following Han et al.~\cite{han2025reevaluating}, we tuned the following: hidden dimensions, attention heads, encoder blocks, dropout rate, and learning rate (shown in Table~\ref{tab:model_tuning}). Batch size (128) balances training time and generalization~\cite{restack_transformer_tuning}. The global model preferred more encoder blocks (4 vs. 2) with identical hidden dimensions and attention heads, reflecting that simpler models generalize better on limited per-vehicle data.

\rev{\textbf{Baseline Tuning.} All baselines were tuned using the same Hyperband setup with equivalent computational budgets. Search spaces were adapted to each architecture's native hyperparameters (e.g., recurrent layers, filter sizes, hidden units) with comparable complexity ranges. All models share the same learning rate, dropout, batch size, and early-stopping patience. The only difference is the loss function: baselines use standard masked BCE with positive weighting, while \textsc{AIMformer} uses \ac{PFBCE}, ensuring performance differences reflect architectural merit rather than tuning advantage.}

\begin{table}[h]
\centering
\footnotesize
\caption{Hyperparameter tuning for both model types.}
\resizebox{\columnwidth}{!}{%
    \begin{tabular}{|l|c|c|c|}
    \hline
    \multirow{2}{*}{\textbf{Hyperparameter}}
     & \multirow{2}{*}{\textbf{Search Space}} & \multicolumn{2}{c|}{\textbf{Value}} \\
    \cline{3-4}
     &  & \textbf{Global} & \textbf{Individual} \\
    \hline
    Hidden Dimension & \{128, 256, 320, 384\} & 128 & 128 \\
    Number of Attention Heads & \{2, 4, 6\} & 2 & 2 \\
    Number of Encoder Blocks  & \{2, 3, 4\} & 4 & 2 \\
    Dropout Rate & [0.1, 0.2] & 0.1 & 0.1 \\
    Learning Rate & [$10^{-5}$, $10^{-3}$] (log sampling) & $1 \times 10^{-4}$ & $1 \times 10^{-4}$ \\
    \hline
    \end{tabular}
}
\label{tab:model_tuning}
\end{table}

\textbf{Loss Tuning.} Based on Eq.~\ref{eq:loss}, we tuned weights and thresholds to balance \acp{FP} and \acp{FN}. Reducing positive weight ($\lambda_{\text{pos}}$) improved learning, while increasing attack threshold (0.5 $\rightarrow$ 0.6) improved classification metrics (Sec.~\ref{sec:metrics}). We investigated an F1-score-based loss; however, \ac{PFBCE} provided superior balance without sacrificing precision (Table~\ref{tab:loss_tuning}), while \rev{recall remained high as attack patterns are distinct in the feature space. The threshold $\tau = 0.6$ and penalty $\lambda_{\text{FP}} = 1.7$ were selected via validation-set grid search, maximizing F1-score while minimizing false alarms in platoon operations. The positive weight $\lambda_{\text{pos}} = 0.6 < 1$ intentionally downweights the attack class, reducing aggressive attack predictions that would inflate \acp{FP}, while $\lambda_{\text{FP}} > 1$ further penalizes residual \acp{FP}.}

\begin{table}[h]
\centering
\caption{Loss function tuning.}
\resizebox{0.7\columnwidth}{!}{%
\footnotesize
    \begin{tabular}{|c|c|c|}
    \hline
    \textbf{Parameters} & \textbf{Search Space} & \textbf{Value} \\
    \hline
    Loss Function & \{\ac{PFBCE}, BCE, F1BCE\} & \ac{PFBCE} \\
    $\lambda_{\text{FP}}$ & \{1.6, 1.7, 1.8, 1.85, 1.9, 2\} & 1.7 \\
    $\lambda_{\text{pos}}$ &  \{0.5, 0.7, 0.75, 0.9, 1\} & 0.6  \\
    $\tau$ & \{0.5, 0.6, 0.7\} & 0.6\\
    \hline
    \end{tabular}%
}
\label{tab:loss_tuning}
\end{table}

All models were trained, tuned, and evaluated using the attack data ratios in Table~\ref{tab:attack_ratio}. The variations reflect controller susceptibility to specific attacks, with vehicles downstream of the attacker (position 2) experiencing higher attack exposure. \rev{To prevent data leakage from overlapping windows, train/validation/test splits are performed at the \emph{scenario} (trace) level: all windows from a given simulation run appear in exactly one split, ensuring no overlapping or near-duplicate windows cross split boundaries despite the 100$ms$ stride producing windows sharing 9 of 10 timesteps. The 100$ms$ stride mirrors the \ac{CAM} dissemination frequency used in simulation; however, since the model consumes only ego-vehicle sensor data, deployment-time inference can operate at the native sensor sampling rate.}

\begin{table}[h]
\centering
\caption{Attack ratio [\%] shown as Train/Val/Test.}
\label{tab:attack_ratio}
\setlength{\tabcolsep}{3pt}
\small
\resizebox{\columnwidth}{!}{%
\begin{tabular}{|c|ccc|ccc|ccc|ccc|}
\hline
\multirow{2}{*}{\textbf{Vehicle}} & \multicolumn{3}{c|}{\textbf{Controller 1}} & \multicolumn{3}{c|}{\textbf{Controller 2}} & \multicolumn{3}{c|}{\textbf{Controller 3}} & \multicolumn{3}{c|}{\textbf{Controller 4}} \\
\cline{2-13}
& Train & Val & Test & Train & Val & Test & Train & Val & Test & Train & Val & Test \\
\hline
Global    & 12.66 & 12.94 & 12.72 & 11.14 & 12.26 & 9.63 & 17.36 & 17.69 & 15.44 & 20.22 & 21.14 & 19.81 \\
Car 1     & 11.02 & 11.09 & 11.26 & 5.00 & 4.94 & 5.04 & 15.44 & 16.57 & 12.70 & 14.40 & 15.72 & 14.59 \\
Car 2     & 11.22 & 11.38 & 11.26 & 6.35 & 6.94 & 5.04 & 15.78 & 16.92 & 13.01 & 14.43 & 15.74 & 14.59 \\
Car 3     & 13.63 & 14.88 & 13.06 & 13.61 & 15.47 & 11.60 & 17.67 & 17.43 & 16.46 & 22.21 & 24.48 & 21.10 \\
Car 4     & 18.53 & 18.77 & 18.94 & 18.17 & 19.88 & 15.70 & 25.62 & 25.32 & 23.10 & 34.77 & 35.13 & 34.32 \\
Car 5     & 18.69 & 18.79 & 18.94 & 19.20 & 21.55 & 15.80 & 25.39 & 25.13 & 22.91 & 33.83 & 34.09 & 33.54 \\
Car 6     & 20.95 & 20.79 & 20.91 & 24.27 & 26.07 & 23.01 & 29.80 & 31.70 & 28.98 & 25.30 & 26.25 & 22.09 \\
\hline
\end{tabular}%
}
\end{table}

%% file: Sections/5_perfromance_new.tex
\section{Performance Evaluation}
\label{sec:performance}

\subsection{Evaluation Metrics}
\label{sec:metrics}

Our testing process encompasses three evaluation types: (i) global model on whole-platoon data (\textit{General Input}), (ii) global model on individual vehicle data (\textit{Vehicle-specific Input}), and (iii) individual models on their own vehicle data. This investigation reveals trade-offs between locally- and edge-deployed models in platooning environments.
We utilize binary classification metrics (i.e., \emph{recall}, \emph{precision}, \emph{$F_1$ score}, \emph{accuracy}). With padding and masking, the underlying \ac{TP}, \ac{FP}, \ac{FN}, \ac{TN} are:

{\footnotesize
\begin{equation}
\begin{aligned}
\mathrm{TP} &= \sum_i m_{i,j}\, \mathbf{1}(y_i = 1 \land \hat{y}_i = 1), \\
\mathrm{FP} &= \sum_i m_{i,j}\, \mathbf{1}(y_i = 0 \land \hat{y}_i = 1), \\
\mathrm{FN} &= \sum_i m_{i,j}\, \mathbf{1}(y_i = 1 \land \hat{y}_i = 0), \\
\mathrm{TN} &= \sum_i m_{i,j}\, \mathbf{1}(y_i = 0 \land \hat{y}_i = 0).
\end{aligned}
\end{equation}
}

For model comparison, we use \ac{ROC} (displaying \ac{TPR} over \ac{FPR}) and \ac{AUC}. We define \acf{PG} for vehicle $v$ and metric $m$ as:

{\footnotesize
\begin{equation}
\mathit{PG}_{v,m} = \mathrm{Individual}_{v,m} - \mathrm{Global}_{v,m},
\label{eq:perf_gain}
\end{equation}
}

where $\mathrm{Individual}_{v,m}$ and $\mathrm{Global}_{v,m}$ represent individual and global model performance, respectively.

\subsection{\textsc{AIMformer} vs. Baseline Architectures}
\label{subsec:model_comparison}

To showcase \textsc{AIMformer} superiority, we implemented seven baseline \ac{DNN} architectures commonly used for time-series and misbehavior detection in \acp{IoV}~\cite{boualouache2023survey, almehdhar2024survey}. Table~\ref{tab:detailed_architecture} summarizes all models, including layers, parameters, and model sizes. \ac{BiLSTM} has the highest parameter count due to bidirectional processing, while unidirectional \ac{LSTM} reduces parameters by 62-63\%. \ac{GRU} uses bidirectional processing with simpler gating ($\approx$ 24\% fewer parameters). Feedforward architectures (\ac{CNN}, \ac{MLP}) have considerably lower parameter counts. The hybrid \ac{CNN}-\ac{LSTM} achieves efficiency through convolutional feature extraction before recurrent processing. The baseline Transformer achieves the smallest sizes (1.6 MB global, 0.4 MB individual; 79\% parameter reduction for individual models). \textsc{AIMformer}'s distinguishing features include: positional encoding with temporal offset awareness (Eqs.~\ref{eq:capital_position} and \ref{eq:global_offset}), three dropout operations per block (0.1 rate), two normalizations per block, and the custom \ac{PFBCE} loss (Eq.~\ref{eq:loss}) that penalizes \acp{FP} while weighting positive samples (Eq.~\ref{eq:pos_weight}). However, this complexity is reflected in the number of parameters (highest) and footprint (12 MB global, 6.2 MB individual). Nonetheless, quantization achieves significant size reduction (Table~\ref{tab:model_sizes}), facilitating edge deployment.

\begin{table*}[!t]
\centering
\caption{Detailed architecture specifications and complexity of baseline models.}
\label{tab:detailed_architecture}
\small
\resizebox{1\textwidth}{!}{%
    \begin{tabular}{|c|l|l|l|p{5cm}|c|c|c|c|}
    \hline
    \multirow{2}{*}{\textbf{Model}} & \multirow{2}{*}{\textbf{Layers}} & \multirow{2}{*}{\textbf{Dense Layers}} & \multirow{2}{*}{\textbf{Dropout}} & \multirow{2}{*}{\textbf{Special Features}} & \multicolumn{2}{c|}{\textbf{Parameters}} & \multicolumn{2}{c|}{\textbf{Model Size [MB]}} \\
    \cline{6-9}
    & & & & & \textbf{Global} & \textbf{Individual} & \textbf{Global} & \textbf{Individual} \\\hline
    LSTM & 3xLSTM(128) & 128, 64, 1 & 0.3, 0.2 & Unidirectional & 379137 & 357633 & 4.5 & 4.2 \\\hline
    BiLSTM & 3xBiLSTM(128) & 128, 64, 1 & 0.3, 0.2 & Bidirectional & 1011969 & 968961 & 12 & 12 \\\hline
    GRU & 3xBiGRU(128) & 128, 64, 1 & 0.3, 0.2 & Bidirectional & 771585 & 739329 & 9 & 8.6 \\\hline
    CNN-LSTM & 2xConv1D + 2xBiLSTM & 128, 64, 1 & 0.3, 0.2 & BatchNorm, MaxPool & 487297 & 479233 & 5.7 & 5.6 \\\hline
    Transformer & MHA(8 heads, 64)+FFN & 128, 64, 1 & 0.1, 0.3, 0.2 & Self-attention, LayerNorm & 129575 & 27179 & 1.6 & 0.417 \\\hline
    CNN & 3xConv1D + GlobalMaxPool & 256, 128, 1 & 0.4, 0.3 & BatchNorm, MaxPool & 233345 & 225281 & 2.8 & 2.7 \\\hline
    MLP & Flatten + 5xDense & 512, 256, 128, 64, 1 & 0.4, 0.3, 0.2, 0.1 & BatchNorm & 427009 & 211969 & 5 & 2.5 \\ \hline\hline
    \textbf{\textsc{AIMformer} (Ours)} & MHA (2 heads, 128) + FFN & 128, 4x or 2x (512, 128), 1 & 0.1 (3x per block) & Positional Encoding, Self-attention, 2xLayerNorm, Residual, Custom Loss & 1057921 & 529537 & 12.3 & 6.2 \\ \hline
    \end{tabular}%
}
\end{table*}

\begin{figure*}[!htbp]
  \centering
  \begin{subfigure}[b]{0.48\linewidth}
    \centering
    \includegraphics[width=\linewidth]{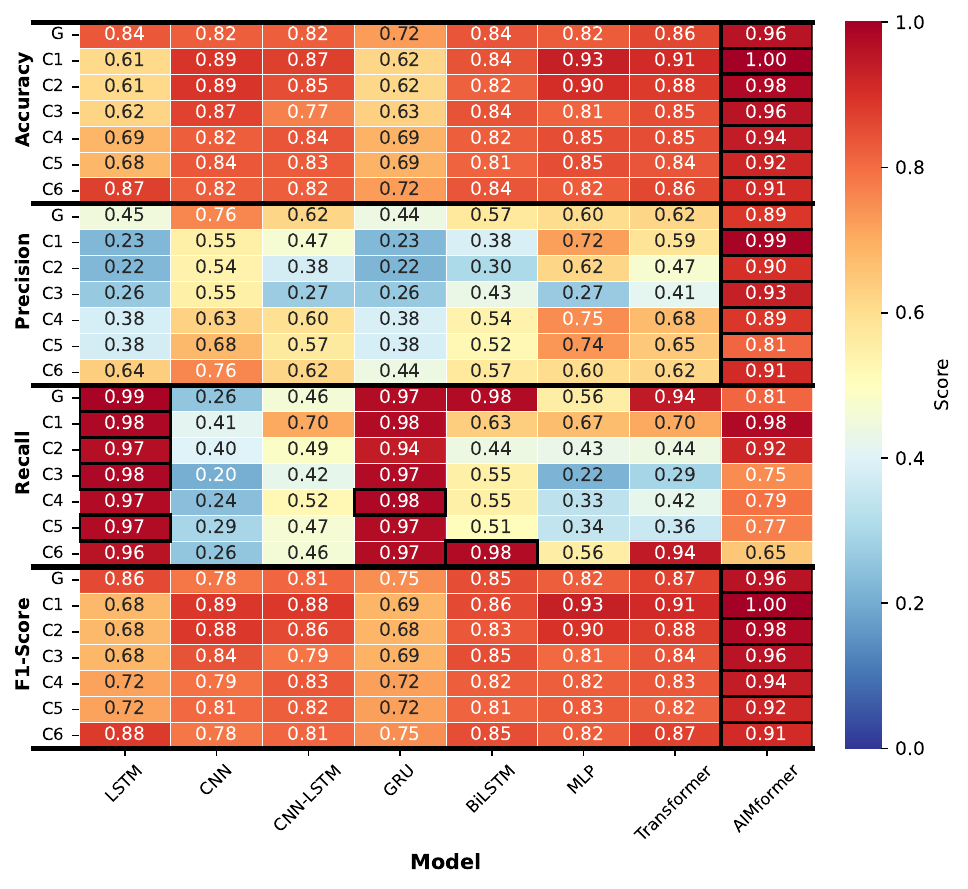}
    \caption{}
    \label{fig:heatmap_c1}
  \end{subfigure}
  \hfill
  \begin{subfigure}[b]{0.48\linewidth}
    \centering
    \includegraphics[width=\linewidth]{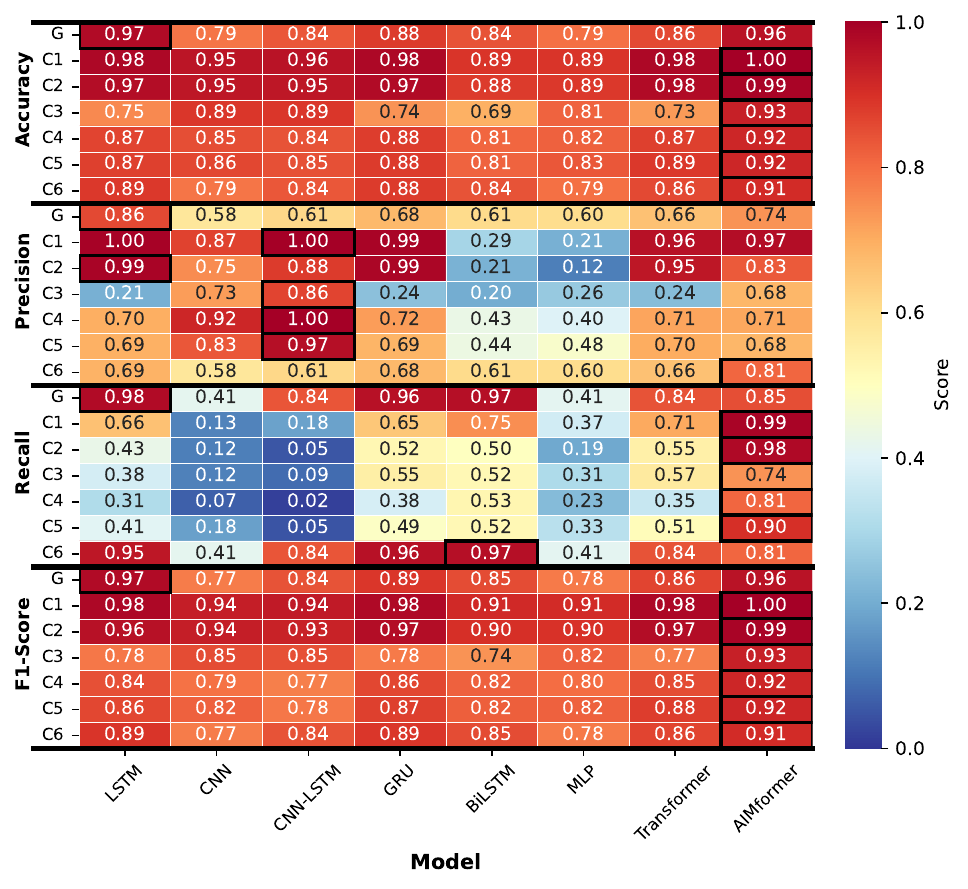}
    \caption{}
    \label{fig:heatmap_c2}
  \end{subfigure}
    
  \begin{subfigure}[b]{0.48\linewidth}
    \centering
    \includegraphics[width=\linewidth]{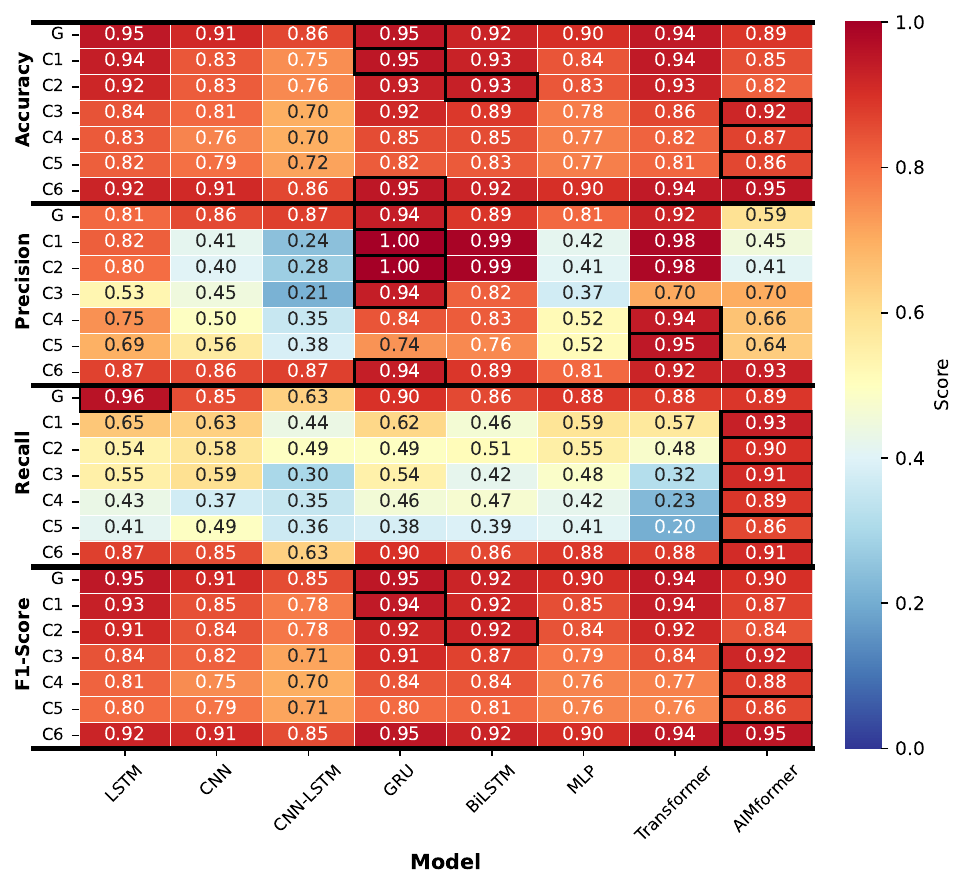}
    \caption{}
    \label{fig:heatmap_c3}
  \end{subfigure}
  \hfill
  \begin{subfigure}[b]{0.48\linewidth}
    \centering
    \includegraphics[width=\linewidth]{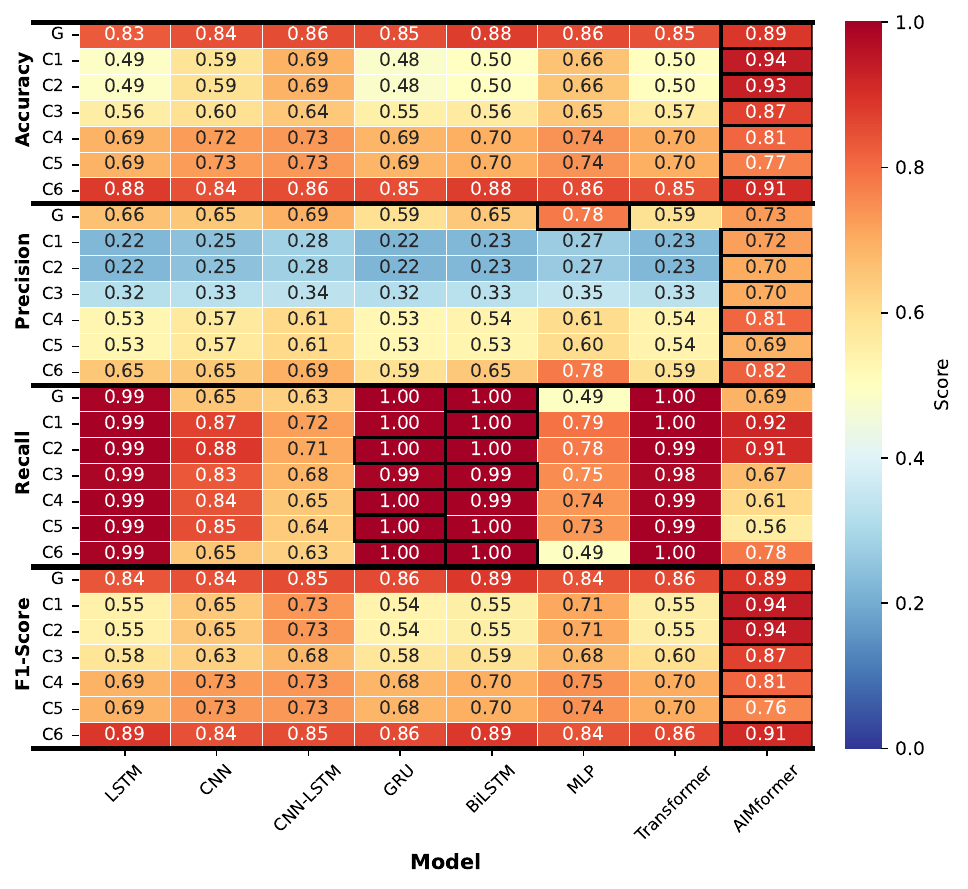}
    \caption{}
    \label{fig:heatmap_c4}
  \end{subfigure}
  
  \caption{Comparison for general and vehicle-specific input. (a) Controller 1, (b) Controller 2, (c) Controller 3, (d) Controller 4.}
  \label{fig:heatmap}
\end{figure*}

Fig.\ref{fig:heatmap} presents classification metrics for all architectures across vehicle positions (C1-C6) and global input (G), for the global model, with black borders indicating the best performers, \rev{averaged over 10 independent training runs
(95\% CI half-widths below $\pm 0.01$ for all cells)}. Across all controllers, \textsc{AIMformer} consistently outperforms baselines. For Controller 1 (Fig.~\ref{fig:heatmap_c1}), most models achieve $>$ 80\% accuracy but struggle with precision-recall balance due to class imbalance; \textsc{AIMformer}'s global input outperforms individual inputs for positions C4-C6. Controller 2 (Fig.~\ref{fig:heatmap_c2}) shows \ac{CNN}-\ac{LSTM} achieving the highest precision but the lowest recall (overly conservative); \textsc{AIMformer} demonstrates superior F1-scores across all positions. For Controller 3 (Fig.~\ref{fig:heatmap_c3}), \ac{GRU} achieves low \acp{FP} but high \acp{FN}; \textsc{AIMformer} maintains high recall (critical for safety) while effectively detecting attacks in challenging positions C3-C5. Controller 4 (Fig.~\ref{fig:heatmap_c4}) validates \textsc{AIMformer}'s balanced precision-recall trade-off, yielding substantially higher F1-scores for all cars.

\begin{figure*}[!htbp]
\centering
    \includegraphics[width=0.7\textwidth]{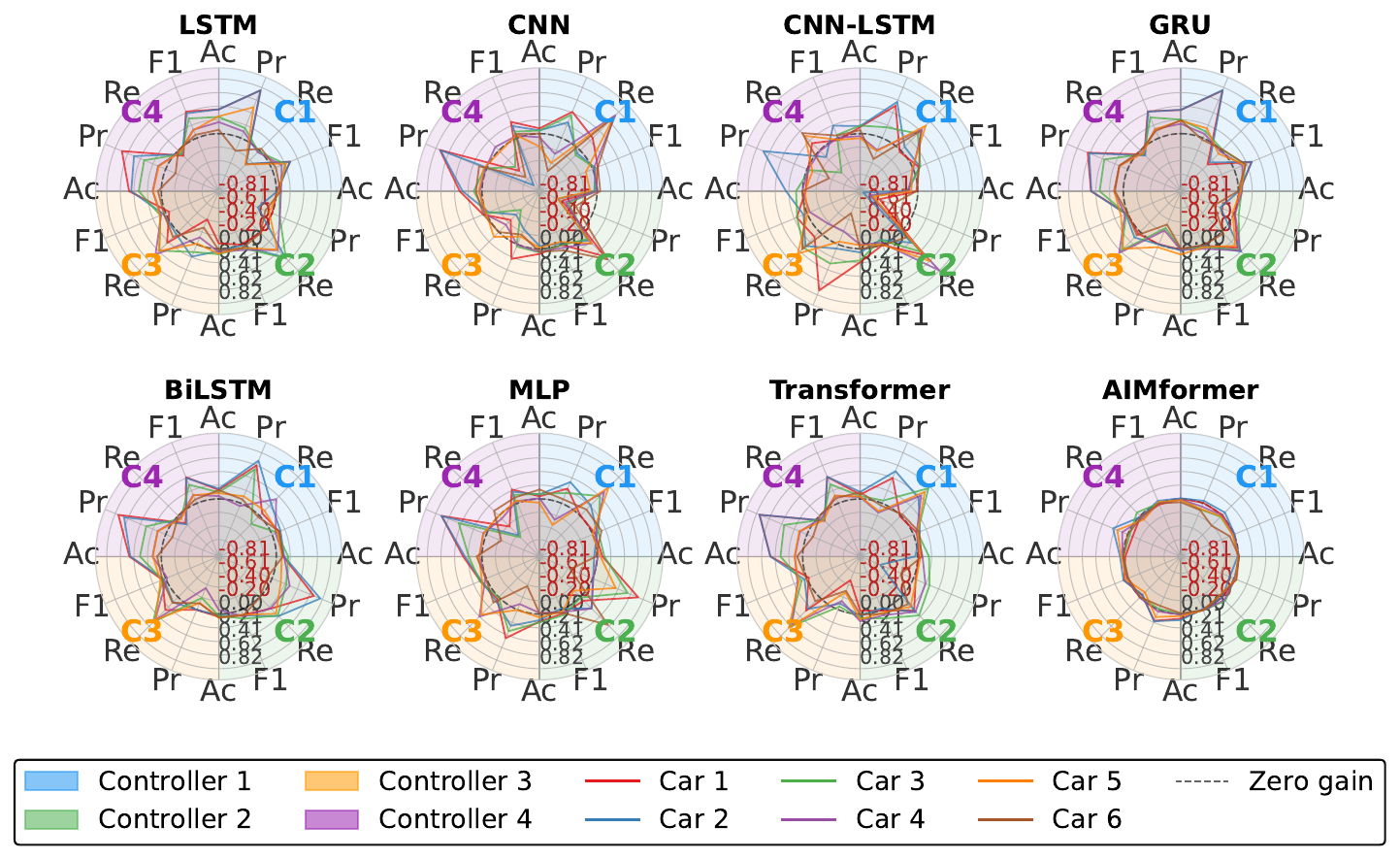}
    \caption{\rev{\ac{PG} controllers comparison. (a) Controller 1, (b) Controller 2, (c) Controller 3, (d) Controller 4.}}
    \label{fig:pf_gain}
\end{figure*}

\rev{Fig.~\ref{fig:pf_gain} illustrates the per-architecture $PG$ (Eq.~\ref{eq:perf_gain}) as radar charts, with each quadrant corresponding to a controller (C1 to C4) and each axis to a classification metric (Ac, Pr, Re, F1). Traces outside the dashed zero-gain circle indicate individual model superiority; traces inside indicate global model superiority. Most architectures exhibit irregular, controller-dependent shapes: \ac{LSTM}, \ac{GRU}, \ac{BiLSTM}, and \ac{MLP} show large outward excursions in C1 and C4 for upstream vehicles (Cars~1 and 2), confirming that individual models consistently outperform the global model at those positions. \ac{CNN} and \ac{CNN}-\ac{LSTM} display the widest per-vehicle spread, with individual cars deviating up to $\pm 0.8$ depending on the controller. In contrast, \textsc{AIMformer} produces near-circular traces tightly around the zero-gain boundary across all four controllers, confirming performance parity between individual and global deployment. This reinforces \textsc{AIMformer}'s suitability for \ac{RSU}-based deployment, where a single global model must generalize across all platoon positions without per-vehicle specialization.}

\subsection{\textsc{AIMformer} Deployment}
\label{subsec:deployment-evaluation}
We evaluate \textsc{AIMformer} on the Jetson Orin Nano Super Developer Kit, employing model quantization~\cite{hussain2024edge} across three optimization frameworks: (i) \ac{TFLite} (direct TensorFlow transformation), (ii) \ac{ONNX} (CPU and CUDA configurations on Jetson Orin Nano Super Developer Kit), and (iii) TensorRT (hardware-optimized). Both individual and global models are optimized using $float16$ and $int8$ quantization.

\begin{figure}[!h]
    \centering
    \includegraphics[width=\linewidth]{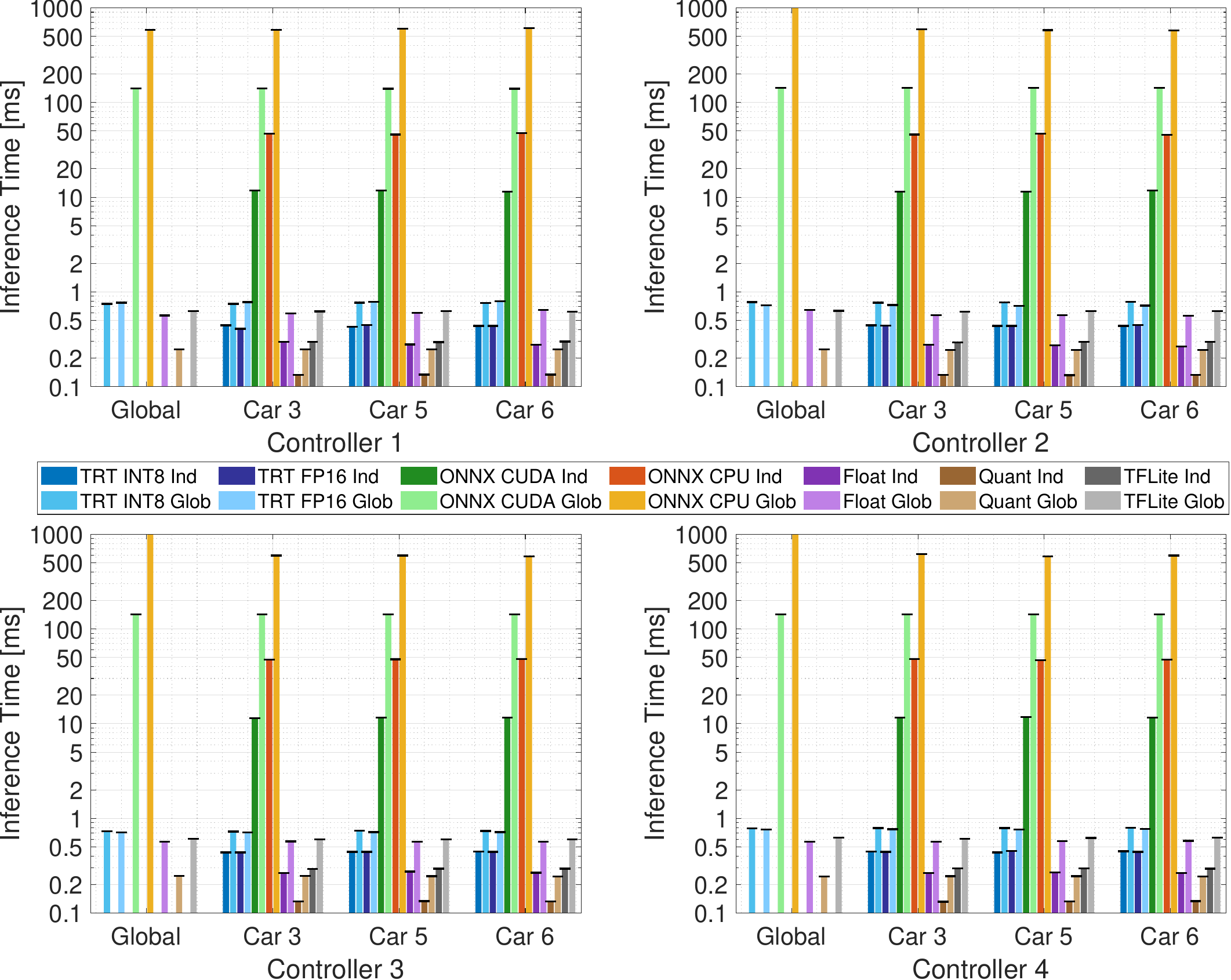}
    \caption{Inference time comparison across quantization methods. Average latency (1000 runs) for global and individual models using \ac{TFLite}, \ac{ONNX}, and TensorRT with float16 and int8 quantization, \rev{with 95\% \ac{CI}.}}
    \label{fig:timing_bar}
\end{figure}

Fig.\ref{fig:timing_bar} presents average inference times for 1000 runs across all controllers. \ac{TFLite} achieves fastest inference (0.13$ms$ for int8 individual models). Inference time differences across controllers are minimal due to similar model inputs and sizes (Table~\ref{tab:detailed_architecture}). Individual models infer 50-90\% faster than global models due to smaller size and input dimensionality. Float quantization is $\approx2.5\times$ slower than integer quantization due to type-casting overhead, indicating a preference for integer quantization without accuracy degradation.

\ac{ONNX} performs significantly worse: CPU-only inference requires $\approx$ 50-600$ms$ (prohibitive for 10Hz \ac{CAM} transmission); CUDA deployment reduces individual model latency to 10$ms$ (viable but suboptimal). TensorRT achieves 0.75-0.8$ms$ (global) and $\approx0.4ms$ (individual), constituting a small fraction of a \ac{CAM} interval, enabling deployment on TensorRT-capable devices. \ac{TFLite} exhibits the fastest inference (0.13$ms$ individual, 0.26$ms$ global), without additional hardware, making it ideal for both in-vehicle and \ac{RSU} deployment.

\rev{\textbf{End-to-End Detection Latency.} \textsc{AIMformer} ($B \cdot V$ configuration) operates on ego-vehicle sensor (e.g., radar, IMU, odometry) and controller output rather than received \ac{V2X} messages, eliminating network-related delays from the detection pipeline. The total detection delay comprises: (i) sensor acquisition, with on-board sensors sampling well above the 10$Hz$ \ac{CAM} periodicity (e.g., automotive radar at 20--50$Hz$, IMU at $\geq$100$Hz$); (ii) look-back buffering ($T=10$ observations); and (iii) model inference (0.13--0.8$ms$ with \ac{TFLite}/TensorRT). After the initial buffer fill, each sensor sample triggers sub-millisecond inference, an order of magnitude faster than the 100$ms$ \ac{CAM} interval bounding network-dependent approaches. Since the system does not depend on \ac{CAM} reception, detection can operate at the native sensor rate rather than the \ac{V2X} dissemination schedule, with the 10 observations look-back horizon remaining the dominant constraint ($\sim$1~second at the training dataset's granularity).}

\begin{figure}[!htbp]
  \centering
  \begin{subfigure}[b]{0.45\linewidth}
    \centering
    \includegraphics[width=\linewidth]{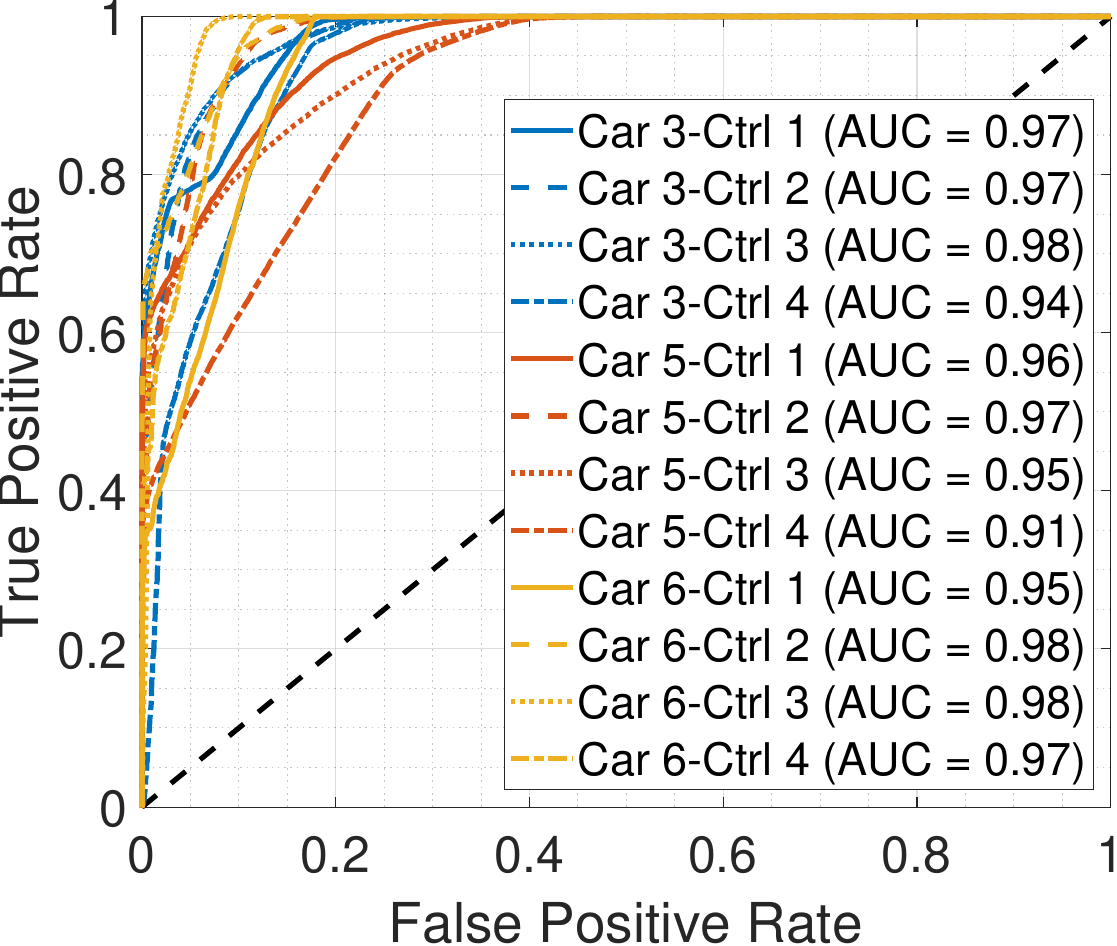}
    \caption{}
    \label{fig:roc-tflite-individual}
  \end{subfigure}
  \hfill
   \begin{subfigure}[b]{0.45\linewidth}
    \centering
    \includegraphics[width=\linewidth]{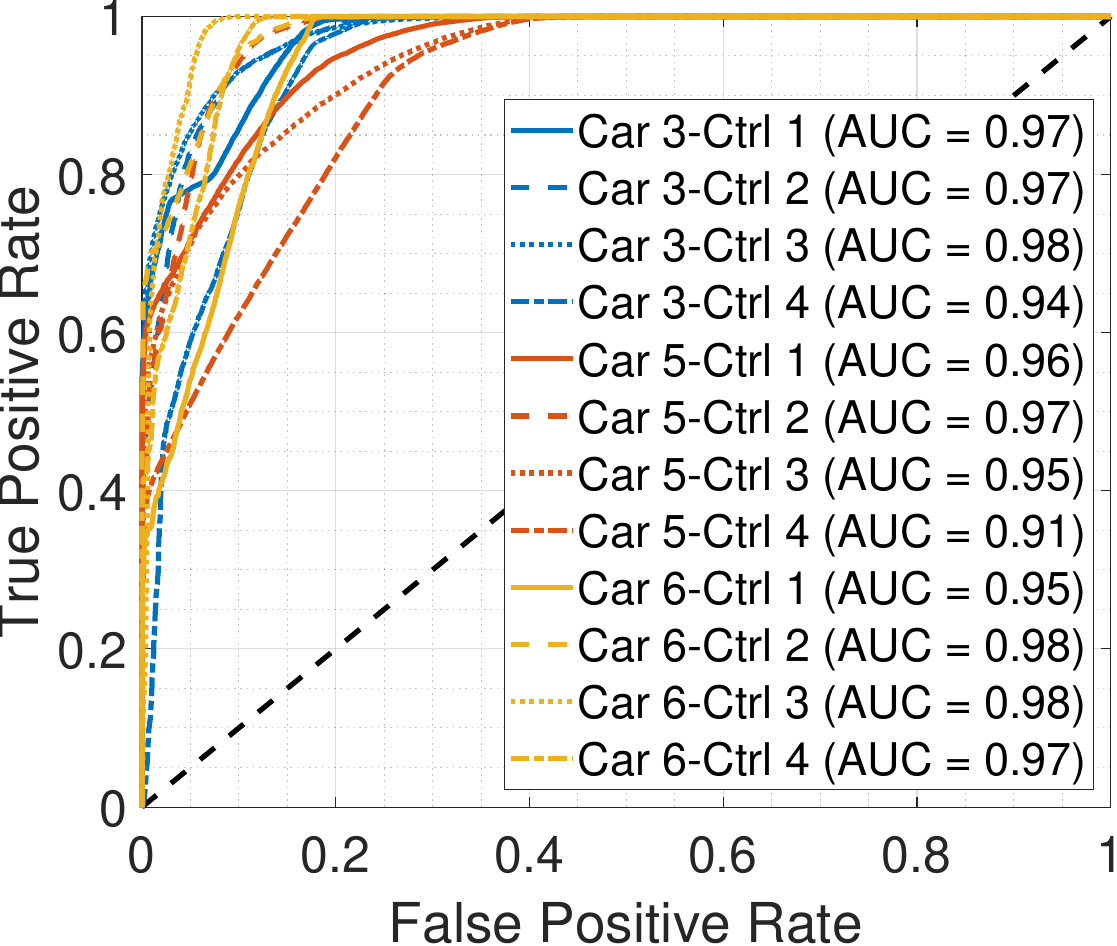}
    \caption{}
    \label{fig:roc-quant-individual}
  \end{subfigure}
  \hfill
  \begin{subfigure}[b]{0.45\linewidth}
    \centering
    \includegraphics[width=\linewidth]{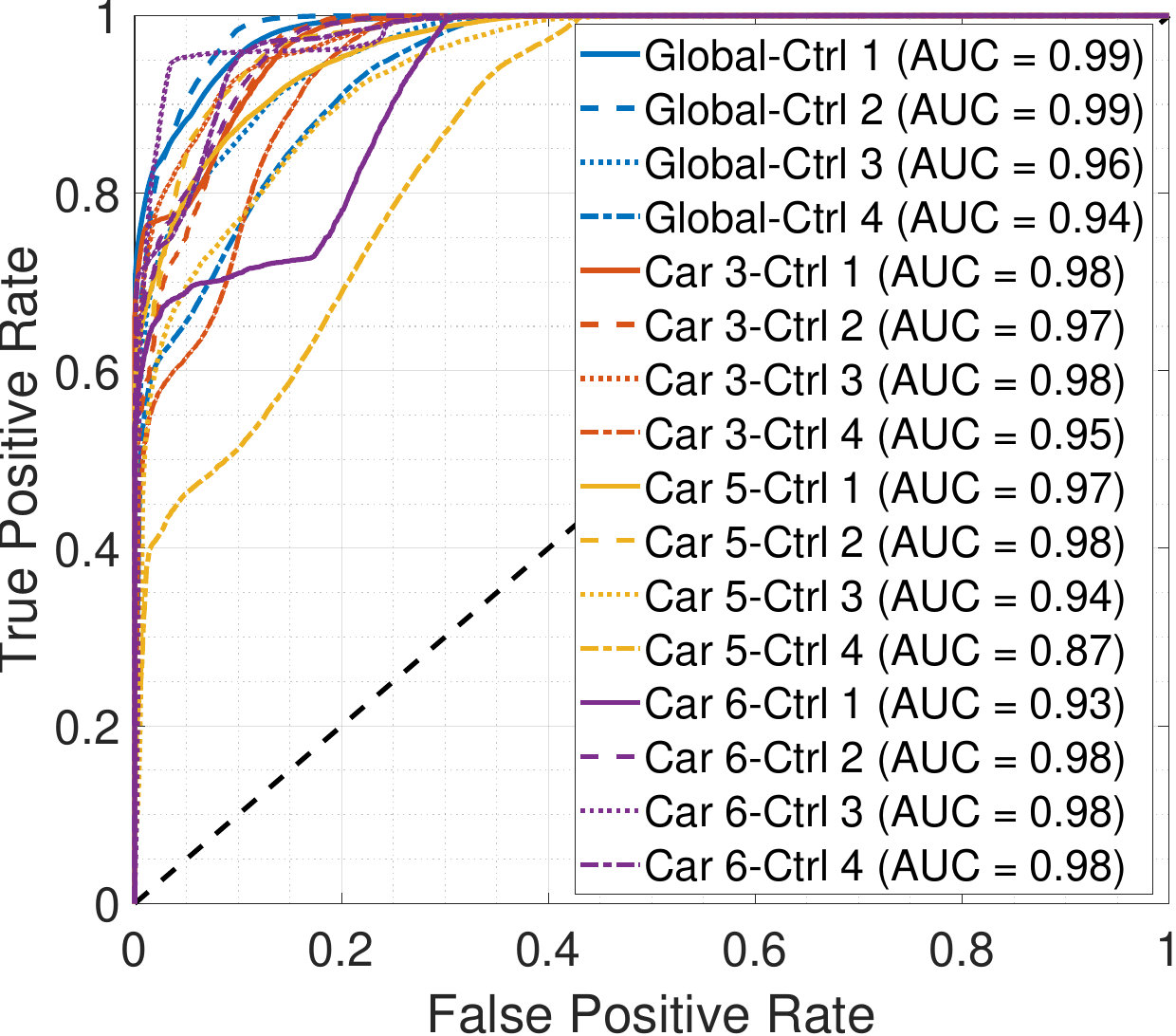}
    \caption{}
    \label{fig:roc-tflite-global}
  \end{subfigure}
  \hfill
  \begin{subfigure}[b]{0.45\linewidth}
    \centering
    \includegraphics[width=\linewidth]{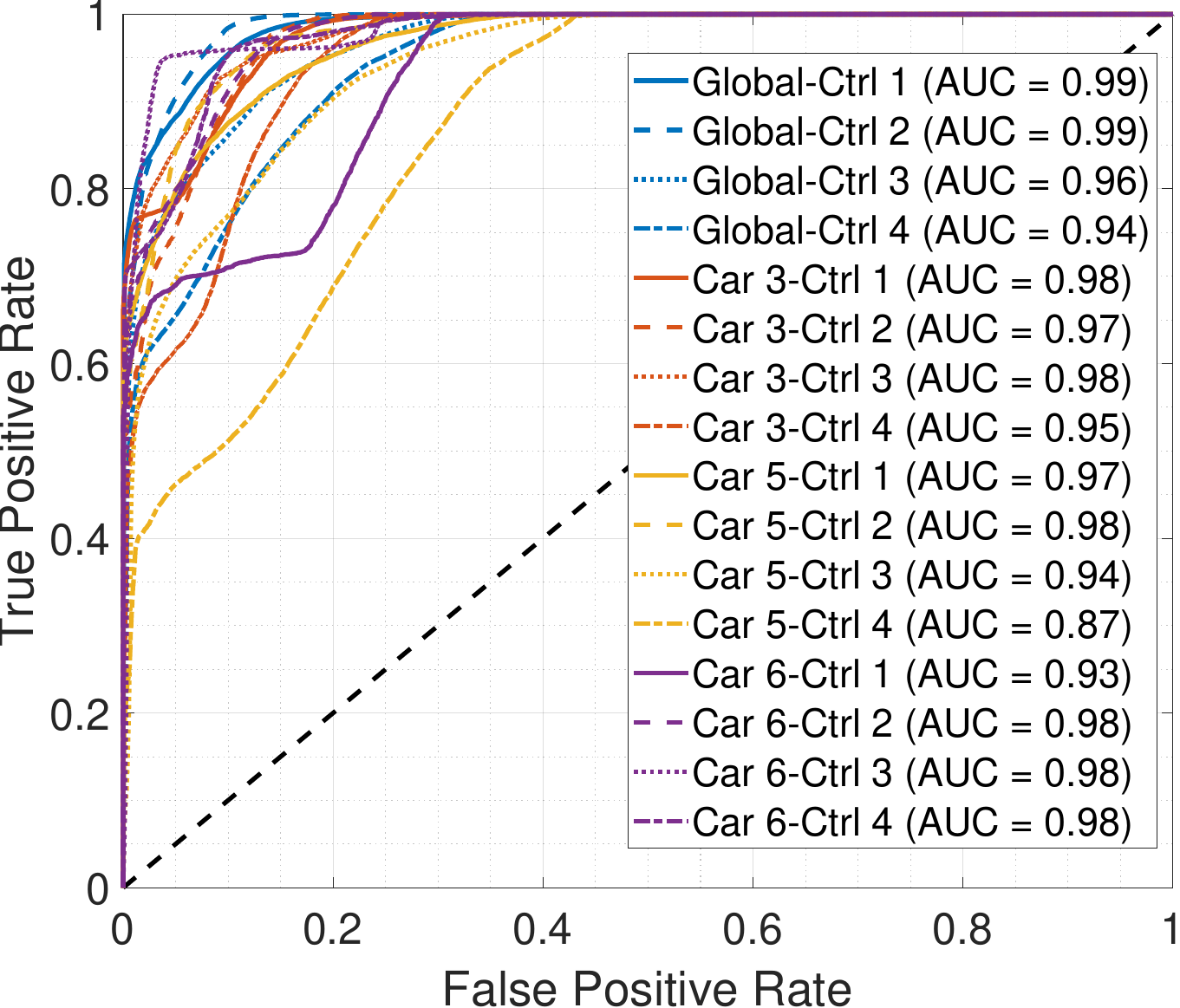}
    \caption{}
    \label{fig:roc-quant-global}
  \end{subfigure}
  \caption{ROC comparison for the \ac{TFLite} and \ac{TFLite} Quantized Individual and Global Models. (a) \ac{TFLite} Individual, (b) Quant Individual, (c) \ac{TFLite} Global, and (d) Quant Global.}
  \label{fig:roc_tflite_quant}
\end{figure}

Fig.~\ref{fig:roc_tflite_quant} compares \ac{TFLite} and quantized model \ac{ROC} curves. Both achieve high \ac{AUC} with nearly identical curves, indicating quantization preserves accuracy. We analyze Cars 3, 5, and 6 (most impactful positions): Car 3 (immediate victim for all maneuvers) achieves high \ac{AUC} with 0.1\% difference between global and individual models; Car 5 (farthest from attacks, most challenging detection position) demonstrates high \ac{AUC}, with global models superior for \ac{CVS} controllers (1 and 2) but individual models superior for \ac{CTH} controllers due to larger intra-platoon distances; Car 6 (joiner) performs similarly to Car 3, with both deployment approaches reliably detecting attacks. The worst performance occurs with the global model under Controller 1 due to divergent lane-change behavior, leading to \acp{FP}, highlighting the limitations of global deployment.

Table~\ref{tab:model_sizes} compares model size. \ac{TFLite} transformation reduces size by 50\% (global) and 27\% (individual); integer quantization achieves 86\% (1.7 MB global) and 80\% (1.2 MB individual) overall reduction. \ac{ONNX} yields similar float quantization sizes (3.4 MB global, 2.3 MB individual) but larger integer sizes (3.3 MB global, 2 MB individual; 94\% and 66\% increases). TensorRT produces larger sizes (4.7 MB float global, 3 MB float/integer individual), with 76-150\% increases over \ac{TFLite}, which also achieves the fastest inference.
\begin{table}[!t]
\centering
\caption{Model sizes comparison.}
\label{tab:model_sizes}
\footnotesize
\begin{tabular}{|>{\centering\arraybackslash}p{2cm}|>{\centering\arraybackslash}p{2cm}|>{\centering\arraybackslash}p{2cm}|}
        \hline
        \multirow{2}{*}{\textbf{Model}} & \multicolumn{2}{c|}{\textbf{Model Size [MB]}} \\
        \cline{2-3}
        & \textbf{Global} & \textbf{Individual} \\\hline
        \textsc{AIMformer} & 12.3 & 6.2 \\ \hline
        \hline
        \multicolumn{3}{|c|}{\textbf{TFLite}} \\\hline
         TFLite & 6.6 & 4.5 \\ \hline
          Float & 3.3 & 2.3 \\\hline
         Dynamic Integer & 1.7 & 1.2 \\\hline\hline
          \multicolumn{3}{|c|}{\textbf{\ac{ONNX}}}\\\hline
         Float& 3.4 & 2.3 \\\hline
          Integer & 3.3 & 2 \\\hline\hline
          \multicolumn{3}{|c|}{\textbf{TensorRT}}\\\hline
          Float & 4.7 & 3 \\ \hline
          Integer & 3 & 3 \\ \hline
    \end{tabular}
\end{table}

\begin{figure}[h!]
  \centering
  \includegraphics[width=0.7\linewidth]{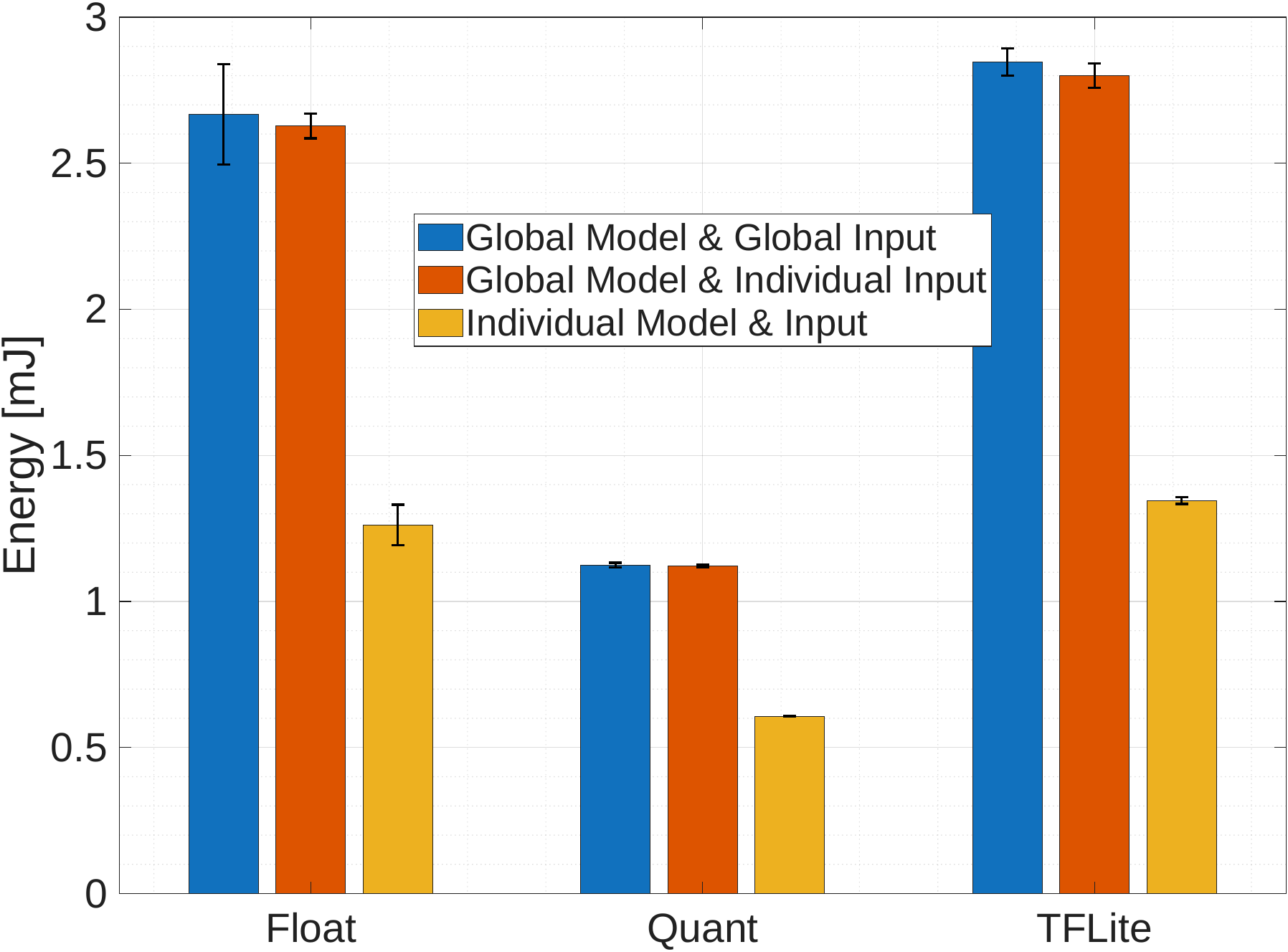}
  \caption{Model energy consumption (1000 runs) with 95\% \ac{CI}.}
  \label{fig:energy_bar}
\end{figure}

\rev{Fig.~\ref{fig:energy_bar} presents the per-inference energy consumption (Jetson 25W mode) across four hardware configurations. Switching from global to individual input leaves global-model energy effectively unchanged ($2.67 \rightarrow 2.63mJ$ for Float, $2.85 \rightarrow 2.80mJ$ for TFLite). Quantized models consistently halve energy ($1.12mJ$ for global, $0.61mJ$ for individual). Individual models reduce energy by 53\% (Float) to 46\% (Quant) relative to the global model. All 95\% CIs remain below $\pm0.17mJ$.}

\subsection{Inter-Vehicle Aware Encoding}
\label{subsec:inter-vehicle-aware-endoding}

The previous sections demonstrated the superiority of the transformer-encoder in detecting misbehavior. Global models outperformed individual ones for downstream vehicles, eliminating the need for multiple models. However, an extra step can be taken by changing Eq.~\ref{eq:reshape} from $\mathbf{X}_{flat} = \text{reshape}(\mathbf{X}, [B \cdot V, T, F])$ to $\mathbf{X}_{flat} = \text{reshape}(\mathbf{X}, [B, V \cdot T, F])$. \rev{In a platoon, controllers enforce speed synchronization across all vehicles; falsification produces cross-vehicle inconsistencies that are strong misbehavior indicators. The $V \cdot T$ reshaping makes cross-vehicle comparisons explicit: self-attention operates over the concatenated $V \cdot T$ dimension, jointly modeling inter-vehicle coordination and intra-vehicle temporal dynamics, while global positional encoding provides temporal synchronization across the platoon.} This requires whole-platoon observations during both training and inference.

\begin{figure}[!htbp]
  \begin{subfigure}[b]{0.48\columnwidth}
    \centering
    \includegraphics[width=\columnwidth]{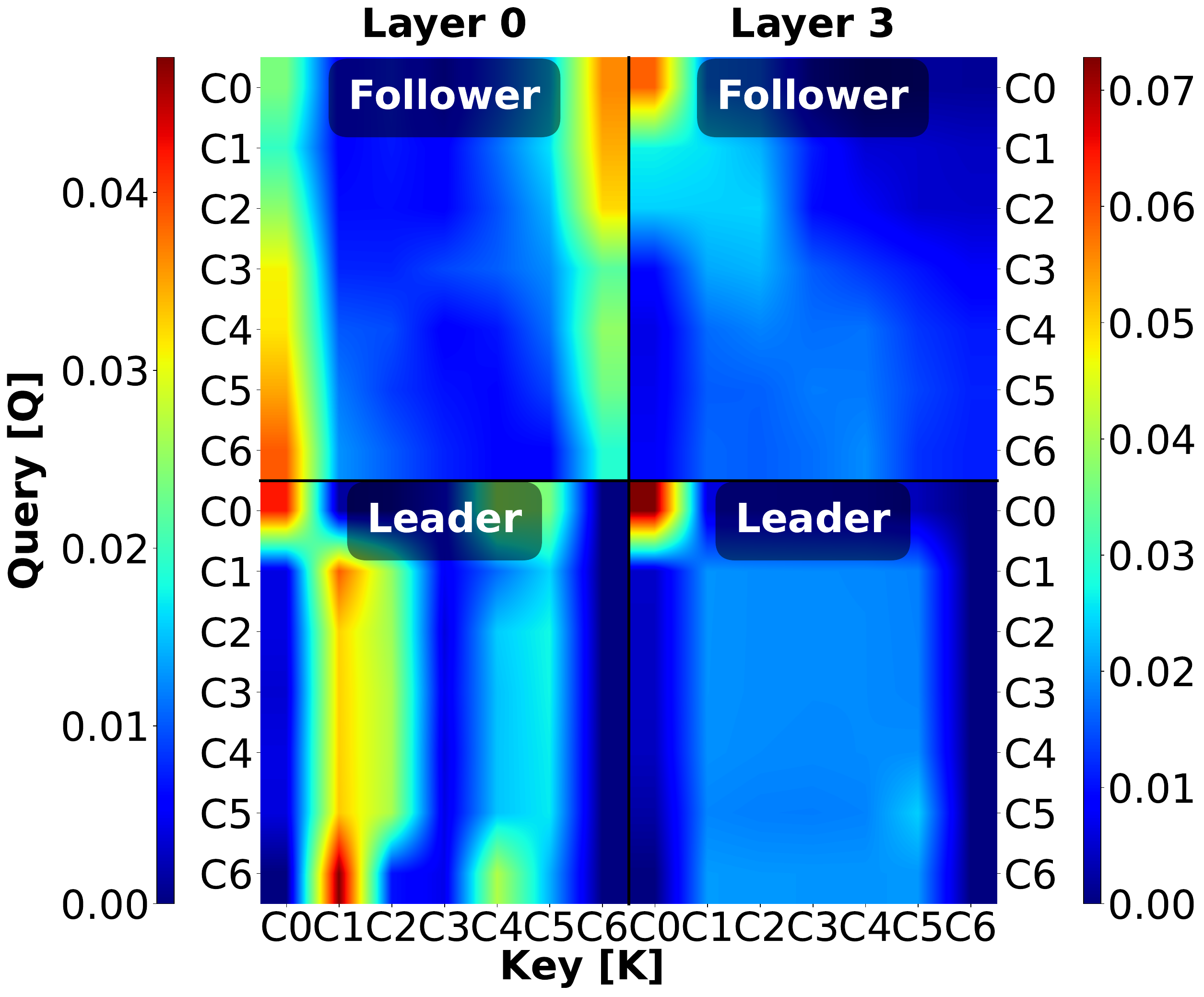}
    \caption{}
    \label{fig:attention_c1}
  \end{subfigure}
  ~  
  \begin{subfigure}[b]{0.48\columnwidth}
    \centering
    \includegraphics[width=\columnwidth]{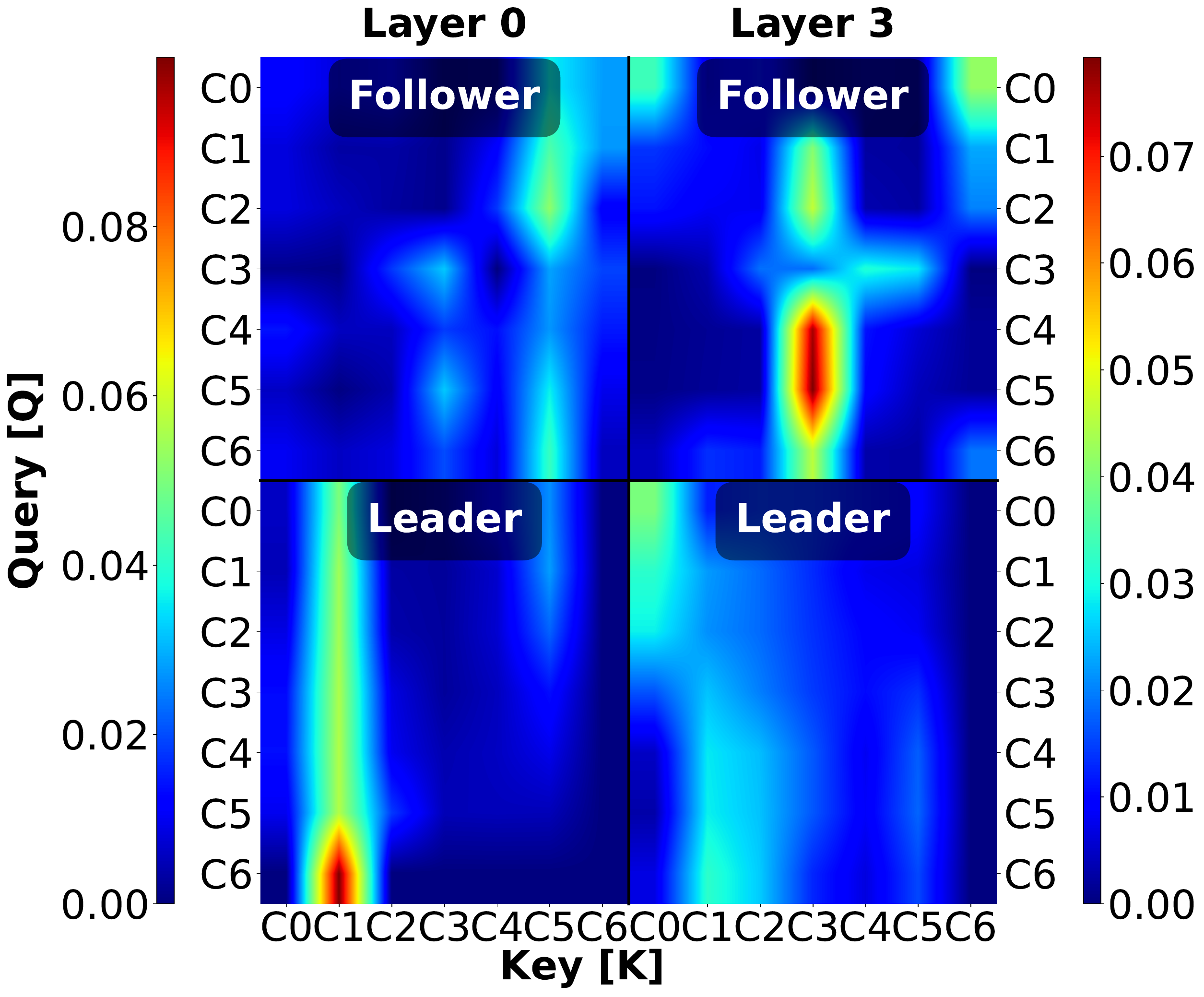}
    \caption{}
    \label{fig:attention_c2}
  \end{subfigure}
  ~
  \begin{subfigure}[b]{0.48\columnwidth}
    \centering
    \includegraphics[width=\columnwidth]{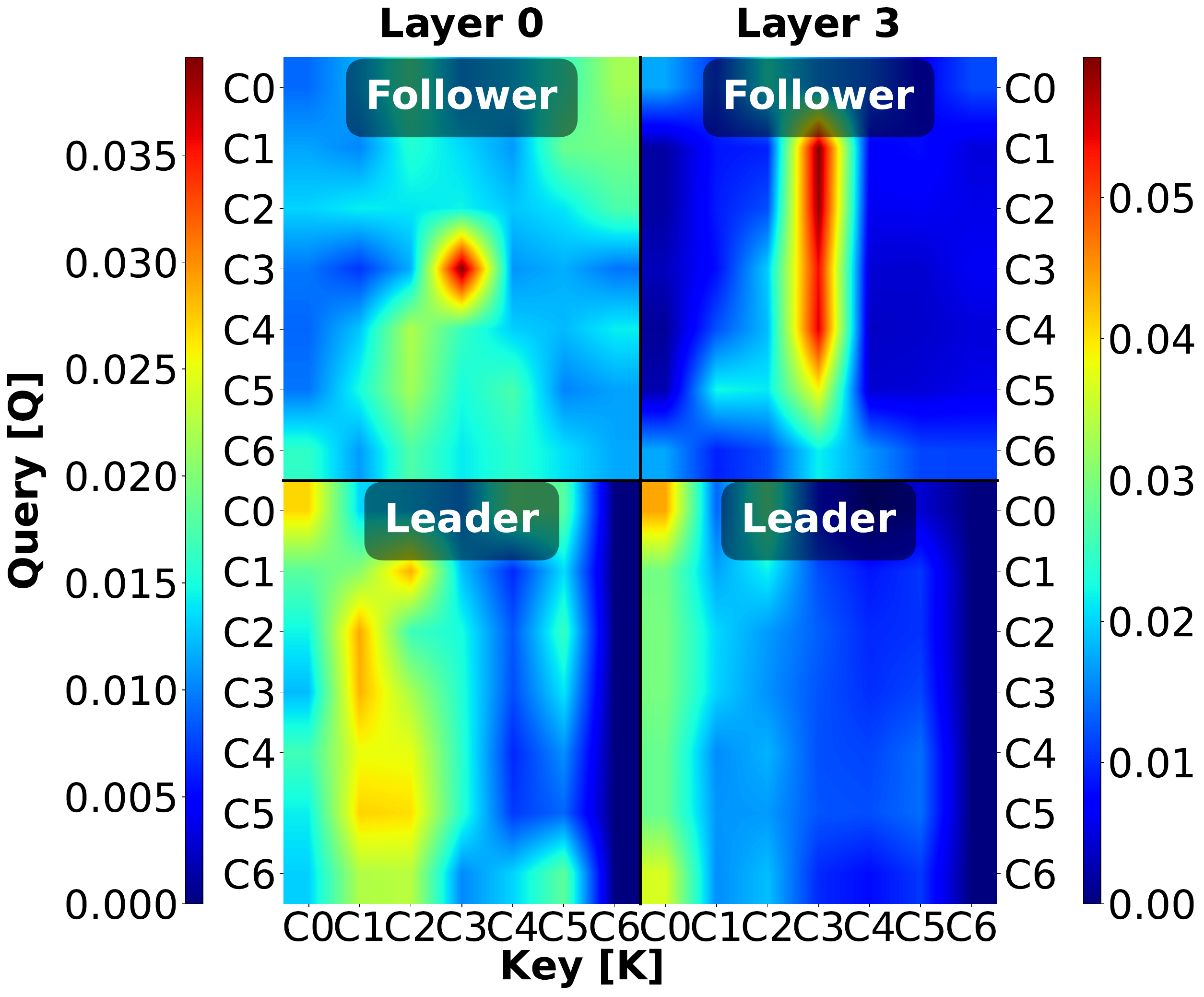}
    \caption{}
    \label{fig:attention_c3}
  \end{subfigure}
  ~
  \begin{subfigure}[b]{0.48\columnwidth}
    \centering
    \includegraphics[width=\columnwidth]{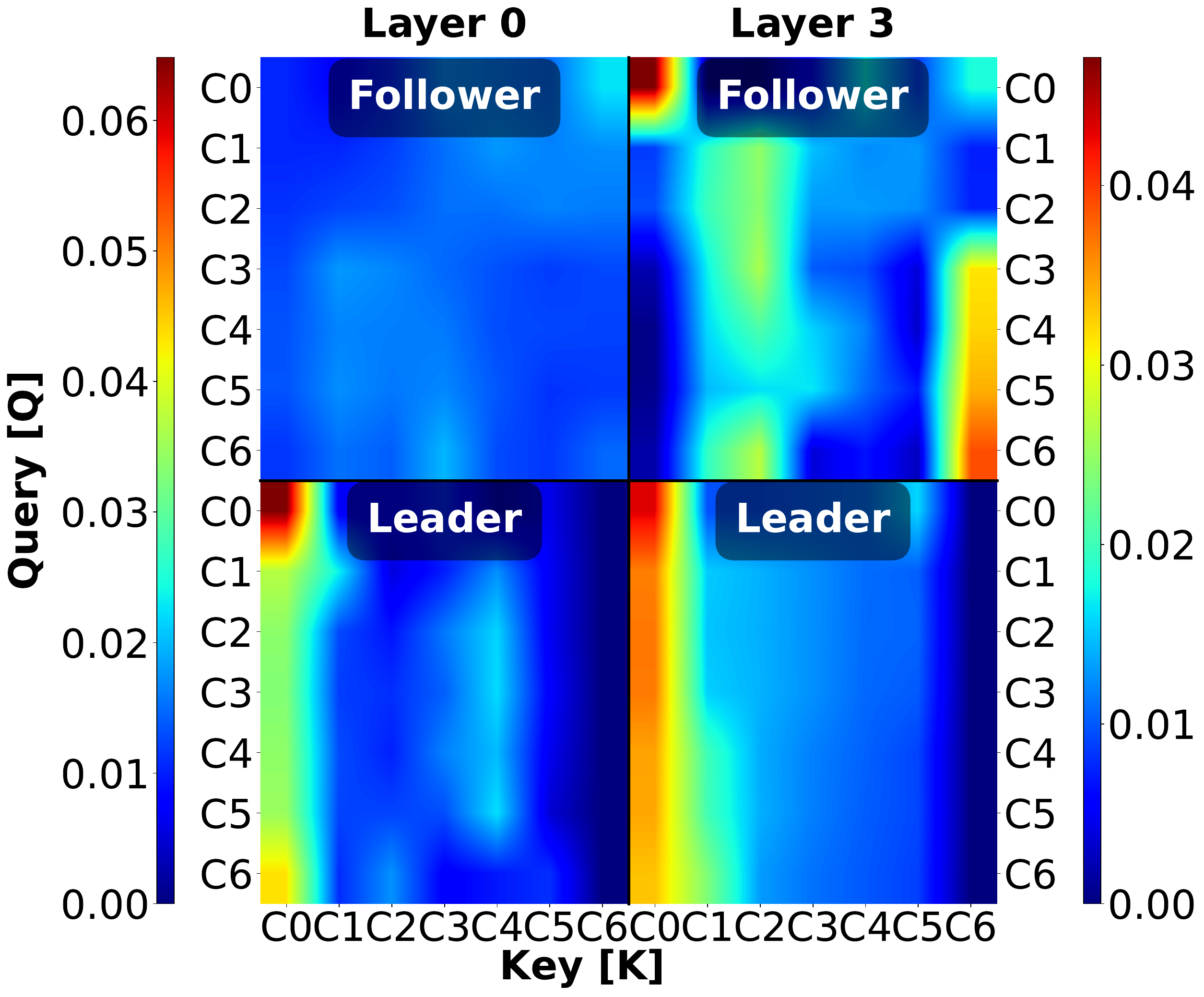}
    \caption{}
    \label{fig:attention_c4}
  \end{subfigure}
  
  \caption{Attention weight controller comparison. (a) Controller 1, (b) Controller 2, (c) Controller 3, (d) Controller 4.}
  \label{fig:attention}
\end{figure}

Fig.~\ref{fig:attention} presents the attention weights for layer 0 (input) and layer 3 (output) across all controllers in the case of attacks. Communication topology fundamentally shapes attention-based security strategies with distinct layer-wise progression and attack-adaptive patterns.

Controller 1 (PATH) employs leader and joiner monitoring for follower attacks (C0-C6 $\rightarrow$ C0, C6 with gradient), while for leader attacks, Layer 0 shifts to victim monitoring (C0-6 $\rightarrow$ C1) and Layer 3 maintains C0-centric patterns, demonstrating attack-adaptive feature extraction with topology-invariant higher layers.
Controller 2 (Ploeg), without leader visibility, progresses from minimal Layer 0 to focused Layer 3 backtracking (C4-6 $\rightarrow$ C3) for follower attacks. For leader attacks, Layer 0 focuses on the victim (C0) while Layer 3 develops position-dependent stratification (C0-2 $\rightarrow$ C0, C3-6 $\rightarrow$ C1), compensating for topology constraints via distributed spatial detection. Controller 3 (Consensus) exhibits a ``victim-versus-attacker'' targeting strategy: follower attacks show C3 $\rightarrow$ C3 self-monitoring (Layer 0) followed by collective validation C0-6 $\rightarrow$ C3 (Layer 3). Meanwhile, the leader attacks unify the followers' attention towards the victim, initially, the attacker, subsequently. Controller 4 (Flatbed) monitors multiple targets for follower attacks: C0 $\rightarrow$ C0, C3-C5 $\rightarrow$ C6, as the joiner integrates behind attacker C2, while simultaneously focusing attention on attacker V2. For leader attacks, it presents a unified C0-C6 $\rightarrow$ C0 attention across both layers.

Notably, this approach enables attack localization through the attention focus. For follower attacks, PATH monitors reference points (C0, C6) without specific attacker localization; Ploeg and Consensus attend to V3 (the immediate victim), enabling indirect localization through predecessor inference; while Flatbed achieves direct attacker (C1-C4 and C6 $\rightarrow$ C2) and victim (C3-C5 $\rightarrow$ C6) identification. For leader attacks, all controllers eventually scrutinize C0, facilitating direct source identification. These findings reveal a spectrum from leader monitoring (PATH) to victim-based inference (Ploeg, Consensus) to explicit source-targeting (Flatbed), with localization capability determined by whether attention focuses on attack symptoms, affected vehicles, or the attacker itself.

\begin{figure}[!htbp]
  \centering
  \begin{subfigure}[b]{0.8\linewidth}
    \centering
    \includegraphics[width=\linewidth]{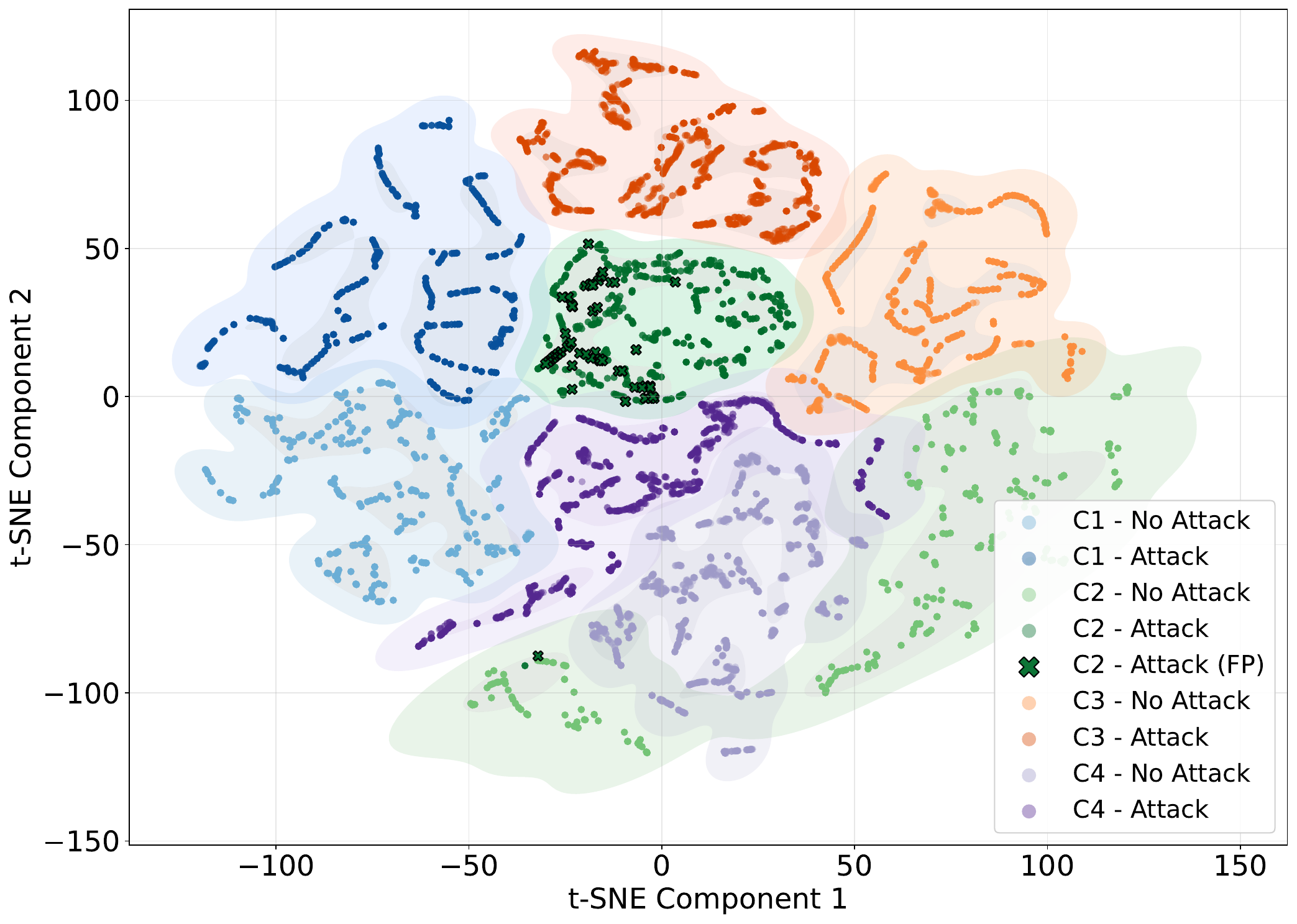}
    \caption{}
    \label{fig:tsne_bv}
  \end{subfigure}
  \begin{subfigure}[b]{0.8\linewidth}
    \centering
    \includegraphics[width=\linewidth]{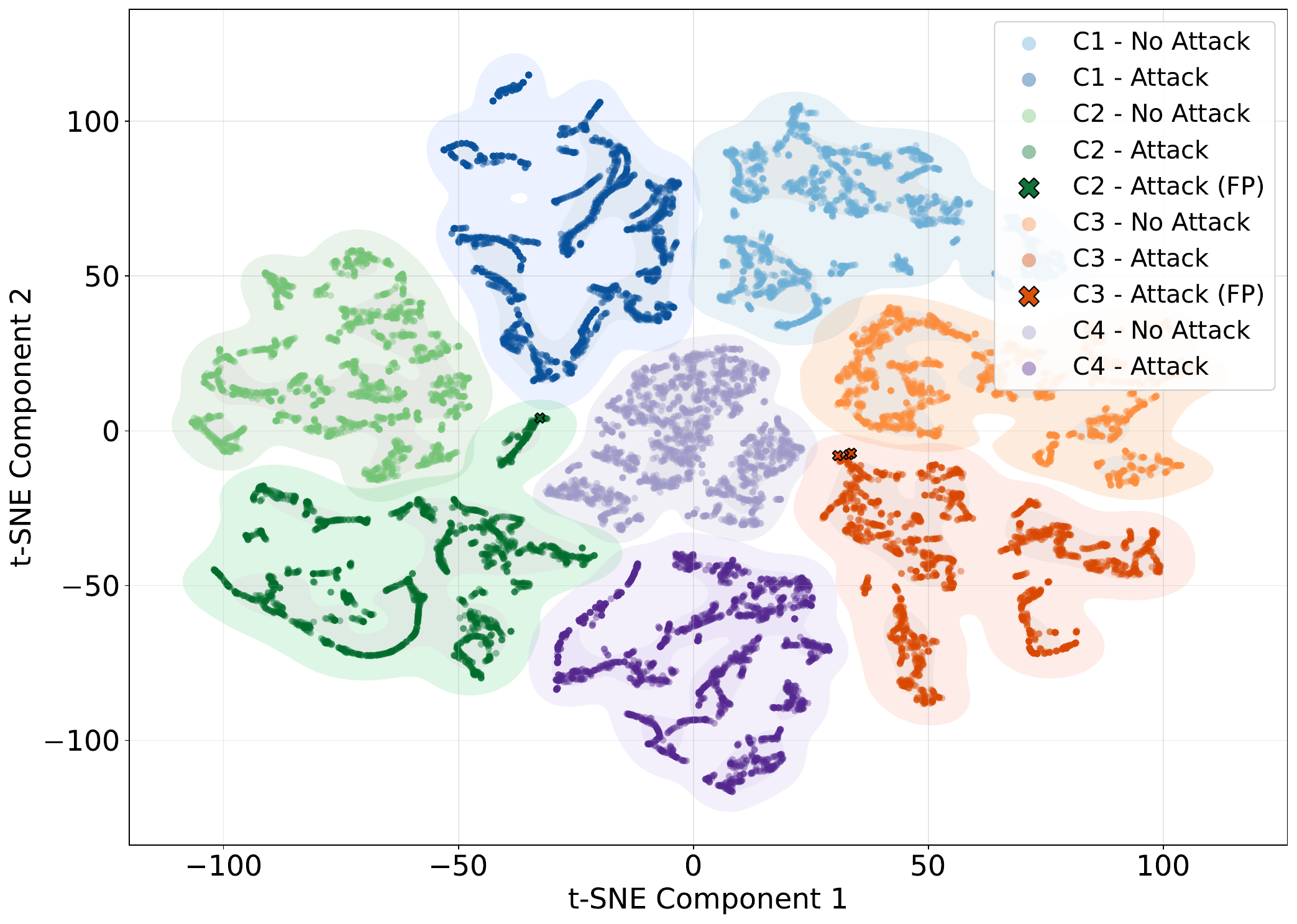}
    \caption{}
    \label{fig:tsne_vt}
  \end{subfigure}
  \caption{t-SNE attention comparison: (a) Attending Time Windows vs. (b) Attending Vehicles.}
  \label{fig:tsne}
\end{figure}

Fig.~\ref{fig:tsne} presents the t-SNE visualizations of the controllers (C1-C4) with learned embeddings from the two transformer-encoder global architectures. The shaded regions are constructed using \ac{KDE}, showing the density regions of the data points. In both cases, we collect the same 2000 samples, allowing for a direct comparison of the clustering. Similar colors represent the classification of benign and attack data points for the same controller, while X colored notations correspond to \acp{FP} for that controller.

Fig.~\ref{fig:tsne_bv} showcases the inference result when the model processes each vehicle's temporal sequences separately, whereas Fig.~\ref{fig:tsne_vt} illustrates the clustering based on the inter-vehicle attention. The former shows overlap between the benign/attack regions of C1 (blue regions), a high number of \acp{FP} for C2, and an overlap between C2 and C4. Conversely, inter-vehicle attention (Fig.~\ref{fig:tsne_vt}) results in each controller occupying distinct regions, with clear separation between benign and attack regions, due to the capture of platoon behavioral patterns. Demonstrated by the \acp{FP} decrease for C2 with 3 \acp{FP} for C3. 

\begin{figure}[!h]
  \centering
  \includegraphics[width=\linewidth]{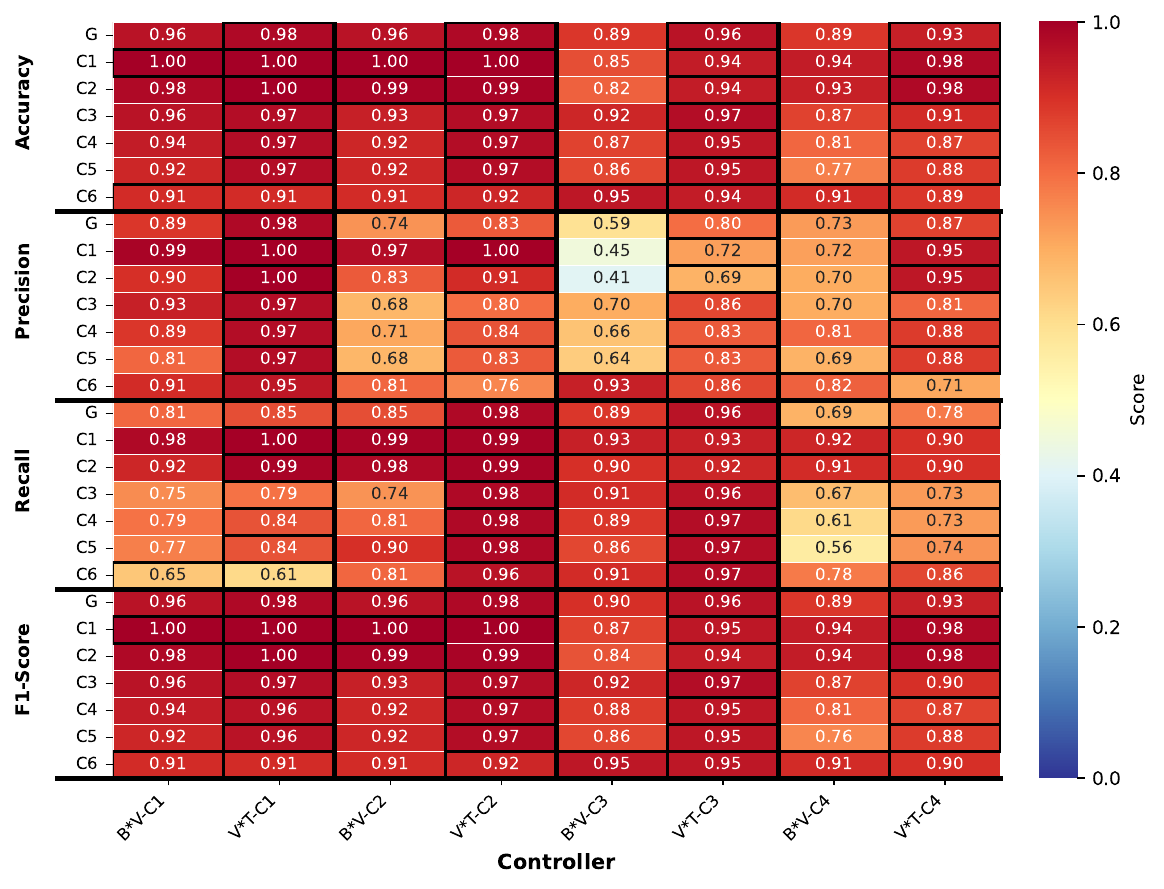}
  \caption{Comparing $B \cdot V$ and $V \cdot T$ global approaches.}
  \label{fig:bv_vs_vt}
\end{figure}

To strengthen the argument for $V \cdot T$'s improvement, we present in Fig.~\ref{fig:bv_vs_vt} the comparison heatmap (similarly to Fig.~\ref{fig:heatmap}). For global inference, $V \cdot T$ consistently outperformed $B \cdot V$, with F1-scores of 0.98, 0.98, 0.96, and 0.93 for controllers C1-C4, respectively, compared to $B \cdot V$'s 0.96, 0.96, 0.90, and 0.89. This advantage persists in individual vehicle inference (keeping in mind that $V \cdot T$ requires whole platoon input), where $V \cdot T$ maintains F1-scores above 0.90 for most vehicles while $B \cdot V$ exhibits greater variability (0.76-1.00).
The performance gap is most evident in precision. $V \cdot T$ achieves near-perfect precision ($\geq$ 0.95) across controllers C1-C4, while $B \cdot V$ shows degradation for specific vehicles (0.41-0.59 for C1 and C2 under controller C3), suggesting that spatio-temporal attention effectively leverages cross-vehicle correlations to reduce false positives. $B \cdot V$ demonstrates competitive recall (0.91-0.93 for controller C3), indicating robust true positive detection despite lacking cross-vehicle context.
The $V \cdot T$ architecture's unified attention over the $V \cdot T$ dimension captures synchronized multi-vehicle patterns, enabling it to leverage learned spatio-temporal priors even during single-vehicle inference. Conversely, $B \cdot V$ learns purely temporal patterns independently per vehicle, trading discriminative power for true vehicle-agnostic deployment capability.
These results reveal a performance-flexibility trade-off. $V \cdot T$ offers superior accuracy (5-10\% higher F1-scores) when complete platoon observations are available, while $B \cdot V$ maintains acceptable performance (F1 $\geq$ 0.87 for almost all cases) with genuine vehicle independence, suitable for distributed deployment where vehicles perform local inference without inter-vehicle communication.

%% file: Sections/6_discussion.tex
\section{Discussion}
\label{sec:discussion}

\textbf{Vehicle-local deployment} (individual models) suits upstream vehicles (positions 1-2) that exhibit position-specific patterns, privacy-constrained scenarios that limit platoon-wide data sharing, heterogeneous platoons with varying capabilities, and bandwidth-limited environments. The 50-90\% inference speedup and 50\% size reduction make individual models optimal for resource-constrained hardware. Nonetheless, a locally deployed $B \cdot V$ global model would enable a single model to be deployed regardless of platoon position, at the cost of increased inference time (still $<1 ms$) and memory footprint.

\textbf{Infrastructure-based deployment} (\ac{RSU}-hosted $V \cdot T$ global models) maximizes detection performance, including attack localization, by capturing platoon-wide input across vehicles and leveraging cross-vehicle correlations. This approach is suitable for scenarios that require consistent platoon-wide performance and high-precision detection to minimize \acp{FP}. \rev{Detection latency is bounded by the \ac{CAM} period: in the worst case, an attacking \ac{CAM} arrives just after the ego vehicle has transmitted, delaying \ac{RSU} detection by up to $\approx100ms$.}

\begin{table}[t]
    \centering
    \footnotesize
    \caption{\rev{V$\cdot$T attack class performance under packet loss.}}
    \label{tab:packet_loss}
    \resizebox{\columnwidth}{!}{%
        \setlength{\tabcolsep}{3pt}
        \begin{tabular}{|>{\centering\arraybackslash}p{1.25cm}|ccc|ccc|ccc|ccc|}
        \hline
       \multirow{2}{*}{\shortstack{\textbf{Packet}\\\textbf{Loss [\%]}}} & \multicolumn{3}{c|}{\textbf{Controller 1}} & \multicolumn{3}{c|}{\textbf{Controller 2}} & \multicolumn{3}{c|}{\textbf{Controller 3}} & \multicolumn{3}{c|}{\textbf{Controller 4}} \\
        \cline{2-13}
         & \textbf{Prec.} & \textbf{Recall} & \textbf{F1} & \textbf{Prec.} & \textbf{Recall} & \textbf{F1} & \textbf{Prec.} & \textbf{Recall} & \textbf{F1} & \textbf{Prec.} & \textbf{Recall} & \textbf{F1} \\
        \hline
        0  & 0.98 & 0.85 & 0.91 & 0.83 & 0.98 & 0.90 & 0.80 & 0.96 & 0.87 & 0.87 & 0.78 & 0.82 \\
        5  & 0.95 & 0.84 & 0.89 & 0.83 & 0.97 & 0.89 & 0.77 & 0.95 & 0.85 & 0.86 & 0.76 & 0.81 \\
        10 & 0.92 & 0.83 & 0.87 & 0.82 & 0.96 & 0.88 & 0.73 & 0.94 & 0.82 & 0.85 & 0.75 & 0.80 \\
        15 & 0.89 & 0.81 & 0.85 & 0.81 & 0.95 & 0.87 & 0.70 & 0.93 & 0.80 & 0.83 & 0.74 & 0.78 \\
        \hline
        \end{tabular}
    }
\end{table}

\rev{To evaluate resilience under packet loss, we randomly drop 5\%, 10\%, and 15\% for vehicle-timestep pairs at inference time, zeroing their features and marking via $\mathbf{M}_{\mathrm{pad}}$. Table~\ref{tab:packet_loss} reports attack-class Precision, Recall, and F1 for the global inference for all controllers. Detection degrades gracefully: at 15\% loss, F1 decreases by 3--7\% per controller, with average drops of 6\%, 4\%, and 5\% in all metrics, respectively. The attention mechanism coupled with $\mathbf{M}_{\mathrm{pad}}$ accommodates missing observations by excluding dropped positions, requiring no architectural modifications or retraining.}

\textbf{Combining} the two methods achieves an optimal balance: vehicles deploy local models for low-latency detection,  while querying \ac{RSU}-hosted global models for accurate detection and localization of the attacker, balancing accuracy, latency, communication overhead, and computational requirements.

\begin{figure}[!h]
\centering
\begin{subfigure}[b]{0.45\linewidth}
  \centering
  \includegraphics[width=\linewidth]{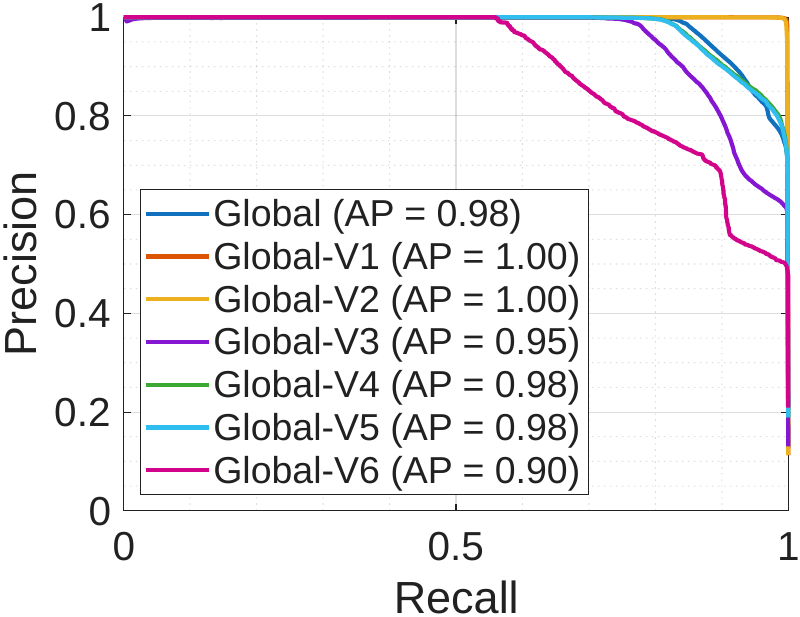}
  \caption{Controller 1}
\end{subfigure}
\begin{subfigure}[b]{0.45\linewidth}
  \centering
  \includegraphics[width=\textwidth]{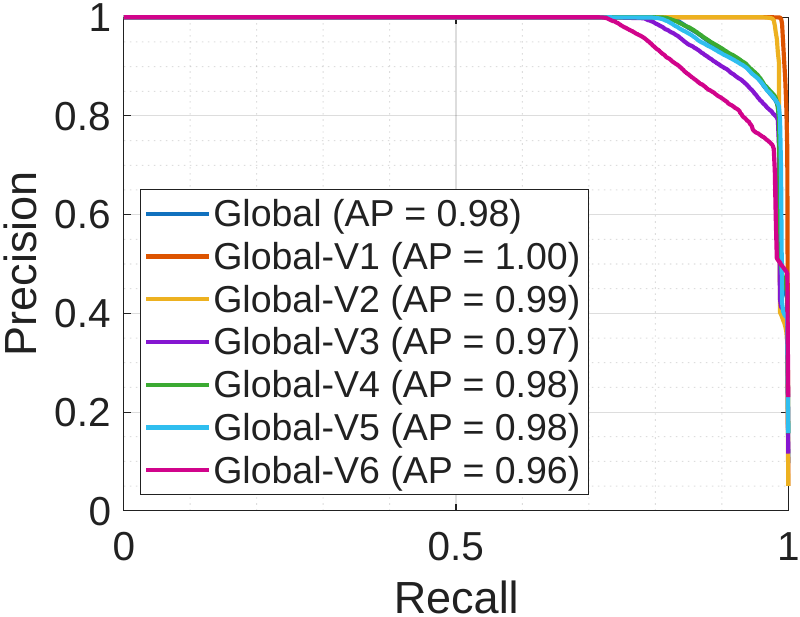}
  \caption{Controller 2}
\end{subfigure}
\begin{subfigure}[b]{0.45\linewidth}
  \centering
  \includegraphics[width=\textwidth]{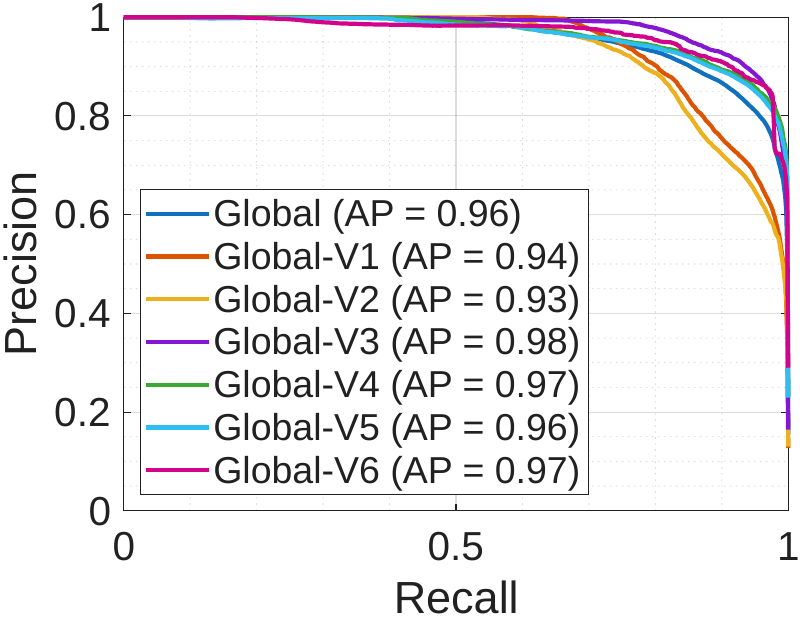}
  \caption{Controller 3}
\end{subfigure}
\begin{subfigure}[b]{0.45\linewidth}
  \centering
  \includegraphics[width=\textwidth]{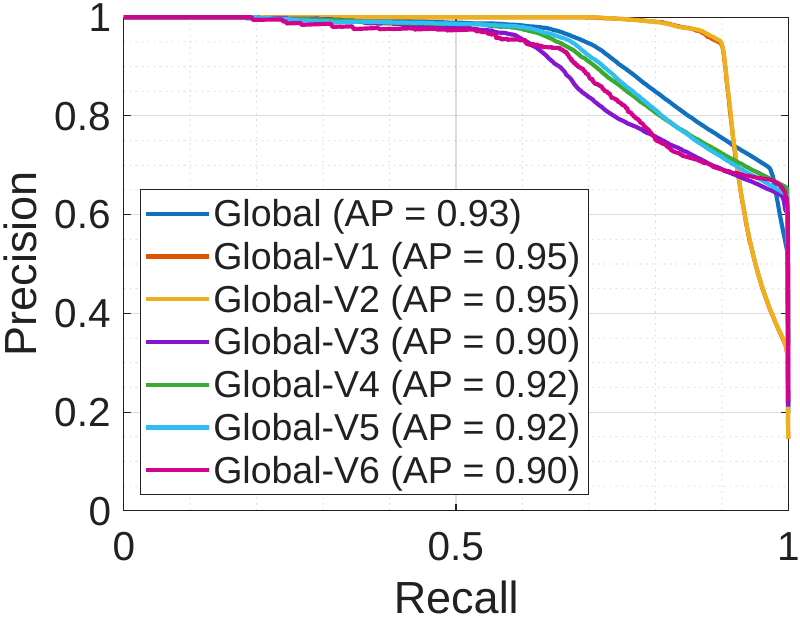}
  \caption{Controller 4}
\end{subfigure}
\caption{\rev{\ac{PR} curves for \textsc{AIMformer} ($V\cdot T$ configuration) across all four platoon controllers, quantifying the \ac{PR} trade-off.}}
\label{fig:pr-curves}
\end{figure}

\rev{\textbf{Threshold Selection and Operating-Point Trade-off.} When $\tau = 0.6$ and $\lambda_{\text{FP}} = 1.7$, the $V \cdot T$ global model achieves mean \ac{TPR} $= 0.89$ (\ac{FNR} $= 0.11$) and \ac{FPR} $= 0.024$ on whole-platoon input. At $10Hz$ this yields approximately one false alarm per $4s$; however, a simple consecutive-positive filter would mitigate this: 2 or 3 consecutive alarms reduce the per-sample \ac{FPR} to $\text{FPR}^{2}$ or $\text{FPR}^{3}$, extending the interval to $\approx 170s$ or $\approx2\,\text{hours}$ respectively, at the cost of $0.2$ or $0.3s$ (for $V \cdot T$ configuration) detection latency. On per-vehicle input, established platoon members (V1--V5) achieve \ac{TPR} $= 0.86$--$0.95$ (\ac{FNR} $= 0.05$--$0.14$) with \ac{FPR} $= 0.016$--$0.039$; the joiner vehicle (V6) achieves \ac{TPR} $= 0.85$ (\ac{FNR} $= 0.15$) and \ac{FPR} $= 0.066$, reflecting the inherently harder detection scenario. The $V \cdot T$ \ac{ROC} \ac{AUC} is 0.95--1.00 across all vehicles and controllers (Fig. omitted due to space), confirming score separation across the full threshold range. The corresponding \ac{PR} curves (Fig.~\ref{fig:pr-curves}) show Average Precision (AP) $\geq0.9$, confirming that precision is maintained for high recall. The $B \cdot V$ global model yields mean \ac{TPR} $= 0.81$ (\ac{FNR} $= 0.19$) and \ac{FPR} $= 0.055$, with \ac{ROC} \ac{AUC} of 0.94--0.99 (Fig.~\ref{fig:roc_tflite_quant}); increasing $\tau$ reduces false alarms at the cost of recall, while lowering it maximizes detection sensitivity.}

\rev{\textbf{Generalization.} \textsc{AIMformer} is evaluated across four heterogeneous controllers spanning distinct communication topologies and spacing policies. Each controller evaluation incorporates three operational scenarios (steady-state, join, and exit) and four platooning speeds, reflecting performance across diverse kinematic conditions. The physics-consistent attacks (``combined'' in Table~\ref{tab:attack_params}), which simultaneously falsify multiple kinematic variables while remaining kinematically feasible and evade simple plausibility checks, are also detected. Since \textsc{AIMformer} operates on controller-induced kinematics rather than raw attack parameters, it detects unknown attacks that produce kinematic disturbances similar to those trained on, providing implicit generalization beyond the trained parameterizations.}

\rev{\textbf{Limitations and Future Directions.} Our simulation-based evaluation (SUMO-Plexe) lacks real-world testbed validation, which remains a future direction alongside adversarial analysis against \textsc{AIMformer}-targeted evasion attacks and \ac{FL} for privacy-preserving deployments.}

%% file: Sections/7_related_work.tex
\section{Related Work}
\label{sec:related_work}

\begin{table*}[!t]
\centering
\caption{Comparison of Misbehavior Detection in Vehicular Networks. Column Deploy: \Circle~= Not appropriate HW (not OBU/Edge); \LEFTcircle~= Appropriate HW (OBU/Edge); \CIRCLE~= HW + Quantization. Metrics: A=Accuracy; P=Precision; R=Recall.}
\label{tab:comparison}
\renewcommand{\arraystretch}{1.25}
\scriptsize
\resizebox{\textwidth}{!}{%
\begin{tabular}{|c|c|p{1.7cm}|c|c|p{1.8cm}|p{1.6cm}|p{1.1cm}||c|c|p{1.7cm}|c|c|p{1.8cm}|p{1.6cm}|p{1.1cm}|}
\hline
\textbf{Ref.} & \textbf{Platoon} & \textbf{Architecture} & \textbf{Deploy} & \textbf{Inference Time [ms]} & \textbf{Metrics [\%]} & \textbf{Attacks} & \textbf{Controller} & \textbf{Ref.} & \textbf{Platoon} & \textbf{Architecture} & \textbf{Deploy} & \textbf{Inference Time [ms]} & \textbf{Metrics} & \textbf{Attacks} & \textbf{Controller} \\
\hline
\hline

\cite{van2018veremi} & No & Plausibility & \Circle & \xmark & P:40-99, R:40-99 & Pos/Speed falsif. & N/A &
\cite{alladi2021securing} & No & 3/4/5x LSTM, CNN-LSTM & \Circle & \xmark & A:98, P:97, R:99 & Pos./spd. falsif. & N/A \\
\hline

\cite{kamel2020simulation} & No & Plaus. + cons., MLP & \Circle & -- & Det:\cmark & Stale msgs., falsif. & -- &
\cite{alladi2023deep} & No & CNN-LSTM, GRU, AE & \Circle & -- & A:98, P:97, R:96 & Pos./spd. falsif. & -- \\
\hline

\cite{so2019physical} & No & RSSI vs. k-NN/SVM & \LEFTcircle & -- & A:87-94, P:87-99, R:61-84 & Pos. falsif. & -- &
\cite{hsieh2021deep_integrated} & No & CNN-LSTM-SVM & \Circle & -- & A:95.4, F1:96.1 & Pos. falsif. & -- \\
\hline

\cite{ercan2022misbehavior} & No & Ens. (RF, SVM, k-NN) & \Circle & -- & A:81-92, P:82-95, R:79-90, F1:\cmark & Pos. falsif. & -- &
\cite{gurjar2025federated_vanet} & No & FedAvg, FedProx & \Circle & -- & A:93, P:92, R:91, F1:92 & 19 VeReMi types & -- \\
\hline

\cite{gyawali2020machine} & No & RF + Dempster-Shafer & \Circle & 2.0 & P:86-99, R:46-99, F1:89-96 & Sybil, DoS & -- &
\cite{kang2016intrusion_dnn} & No & DNN with DBN & \Circle & 2-5 & A:99.7, P:99, R:99, AUC:>99 & Pkt. inj., manip. & -- \\
\hline

\cite{uprety2021privacy} & No & FedML (SVM, LSTM) & \Circle & -- & A:92, P:93, R:90 & Pos. falsif. & -- &
\cite{song2020invehicle_cnn} & No & Deep CNN (Inc-ResNet) & \LEFTcircle & 5-7 & A:99.99, P:99, R:99 & DoS, spoofing & -- \\
\hline

\cite{alladi2021deep} & No & DNN w/ seq. recon. & \Circle & -- & Anom:\cmark, Gain:\cmark & Pos./spd. falsif. & -- &
\cite{longari2021cannolo} & No & LSTM Autoencoder & \LEFTcircle & 650 & AUC:0.97, F1:0.95 & Replay, manip. & -- \\
\hline

\cite{gyawali2019misbehavior} & No & k-NN, RF, Ensemble & \Circle & -- & P:94-98, R:85-93, F1:89-96 & Pos. falsif., DoS & -- &
\cite{javed2021canintelliids} & No & CNN + Attn-GRU & \Circle & -- & A:99, Gain:10.8\% & Mixed intrusion & -- \\
\hline

\cite{kushardianto2021step} & No & DBN, LSTM, GRU, RF & \Circle & 0.05-1.16 & A:75-92 (2-step) & Pos. falsif. & -- &
\cite{sun2021anomaly_cnnlstm} & No & CNN-LSTM + Attn & \Circle & -- & F1:0.95, ER:2.16\%, A:97.8 & Multi-attack CAN & -- \\
\hline

\cite{zhu2019mobile_edge_lstm} & No & Multi-task LSTM & \LEFTcircle & 0.61-144 & A:90, P:89, R:88 & Multi-dim. anom. & -- &
\cite{sarker2018data} & Part. & GMM-MDN w/ RNN & \Circle & -- & Cons:\cmark, Det:\cmark & Vel., acc. falsif. & -- \\
\hline

\cite{kalogiannis2022attack} & \textbf{Yes} & GMM-HMM & \Circle & 130-280 & F1:87, P:88, R:86 & Pos./spd./acc., jam & SUMO CACC &
\cite{boddupalli2021replace} & \textbf{Yes} & RF & \Circle & 10 & Coll:58$\to$2\%, Det:\cmark & Pos. mutation & Santini \\
\hline

\cite{wang2021deep} & Part. & LSTM/GRU & \Circle & -- & Stealth:\cmark, Det:\cmark & Stealth attacks & ACC/CACC &
\cite{ko2021approach} & \textbf{Yes} & LSTM & \Circle & -- & A:$>$96, Det:\cmark & Corr./non-corr. & CACC \\
\hline

\cite{boddupalli2019redem} & Part. & MLP & \LEFTcircle & -- & Det:\cmark, Resil:\cmark & Comm. DoS V2V/V2I & CACC &
\cite{mousavinejad2022secure} & \textbf{Yes} & Set-membership & \Circle & -- & Stab:\cmark, Det:\cmark & DoS, deception & Leader-fol. \\
\hline
\hline

\textbf{Ours} & \textbf{Yes} & \textbf{Transformer (MHA)} & \textbf{\CIRCLE} & \textbf{0.13-0.8} & \textbf{A: \cmark, P: \cmark, R: \cmark, F1: \cmark, AUC:96-99, ROC: \cmark} & \textbf{Pos./spd./acc. (stealth)} & \textbf{SUMO CACC} &
\cite{li2025attentionguard} & \textbf{Yes} & Transformer (MHA) & \Circle & 100-500 & A:88-92, P:91-93, R:88-92, F1:89-92, AUC:96-99 & Pos./spd./acc. (stealth) & Multiple CACC \\
\hline

\end{tabular}%
}
\vspace{2mm}

\end{table*}

This section discusses misbehavior detection in \acp{VC} systems, transformer architectures for security applications, \ac{DL} for vehicular security, and Edge \ac{AI} deployment.

\subsection{Misbehavior Detection in Vehicular Networks}

Misbehavior detection in \ac{IoV} environments has evolved from rule-based plausibility checks to \ac{ML}-based approaches. Van der Heijden et al.~\cite{vanderheijden2019survey} provided a comprehensive survey identifying fundamental challenges in cooperative \ac{ITS}, including the difficulty of distinguishing between benign sensor errors and malicious data falsification. Boualouache and Engel~\cite{boualouache2023survey} extended this analysis to 5G-enabled vehicular networks, emphasizing the need for real-time detection capabilities that scale with increasing network density.

Frameworks have been essential for advancing the field. Kamel et al.~\cite{van2018veremi} introduced the VeReMi dataset, providing the first reproducible benchmark for comparing misbehavior detection algorithms across diverse attack scenarios. The VeReMi Extension~\cite{kamel2020veremi_ext} expanded coverage to include additional attack vectors and environmental conditions, becoming the de facto standard for evaluating detection approaches. So et al.~\cite{so2019physical} demonstrated that combining \ac{PHY}-layer plausibility checks with traditional position verification improves detection rates by 15-20\%, achieving 83-95\% accuracy against falsification attacks.

\ac{ML} integration has shown substantial promise. Sharma and Liu~\cite{sharma2021machine} developed a data-centric model achieving 94\% detection accuracy by leveraging ensemble methods, while Gyawali et al.~\cite{gyawali2020machine} incorporated reputation techniques to reduce \ac{FP} rates in dynamic network topologies. Privacy-preserving approaches have gained attention, with Uprety et al.~\cite{uprety2021privacy} proposing \ac{FL} for distributed misbehavior detection that maintains 92\% accuracy while preserving location privacy.

Platoon security research remains limited. Previous work~\cite{kalogiannis2022attack} characterized attack impact across different platoon controllers and topologies, revealing that acceleration falsification poses the greatest threat to string stability. However, existing detection mechanisms struggle with combined attacks that manipulate multiple kinematic parameters in physically consistent patterns---a gap our transformer-based approach addresses.

\subsection{Transformers for Security and Anomaly Detection}

Transformer architectures revolutionized sequence modeling tasks, with recent applications for cybersecurity. Xu et al.~\cite{xu2022anomaly} introduced \textit{Anomaly Transformer}, leveraging association discrepancy to identify temporal anomalies in multivariate time series with 98.2\% \ac{AUC} on benchmark datasets. Tuli et al.~\cite{tuli2022tranad} proposed \textit{TranAD}, demonstrating that deep transformer networks outperform \ac{LSTM}-based approaches by 8-12\% for multivariate time series anomaly detection through parallel processing of temporal dependencies. 

Self-supervised learning has enhanced the effectiveness of transformers for security tasks. Jeong et al.~\cite{jeong2023anomalybert} developed \textit{AnomalyBERT}, which employs data degradation schemes during pre-training to achieve robust anomaly detection without extensive labeled data, particularly relevant for vehicular environments where attack samples are scarce. Long et al.~\cite{long2024transformer_ids} applied transformers to cloud network intrusion detection, achieving 97.3\% accuracy with significantly reduced false positive rates compared to traditional \ac{DNN} approaches.

Despite these advances, transformer applications for vehicular misbehavior detection remained unexplored. Existing work focuses on network traffic analysis or generic time series anomaly detection, lacking the domain-specific considerations necessary for understanding the dynamics of platoon coordination. Our \textsc{AIMformer} architecture addresses this gap through global positional encoding that captures cross-vehicle temporal relationships and custom loss functions optimized for the requirements of safety-critical vehicular applications.

\subsection{Deep Learning for Vehicular Security}

\ac{DL} approaches have become predominant in vehicular security research, with \ac{LSTM} and \ac{CNN} architectures achieving substantial detection performance. For misbehavior detection in \ac{VANET} environments, Alladi et al.~\cite{alladi2023deep} developed a \ac{DL}-based classification scheme achieving 95.7\% accuracy across nine attack categories in cooperative \ac{ITS}. Hsieh et al.~\cite{hsieh2021deep_integrated} proposed an integrated \ac{CNN}-\ac{LSTM} architecture that captures both spatial feature patterns and temporal dependencies, demonstrating 93.2\% detection rates on the VeReMi dataset. 

In-vehicle network security has received considerable attention. Kang and Kang~\cite{kang2016intrusion_dnn} applied \ac{DNN} for \ac{CAN} bus intrusion detection, achieving 99.7\% accuracy for binary classification tasks. Song et al.~\cite{song2020invehicle_cnn} extended this work with convolutional architectures that automatically extract message timing patterns, while Longari et al.~\cite{longari2021cannolo} employed \ac{LSTM} autoencoders for unsupervised anomaly detection with 99.2\% \ac{AUC}. Javed et al.~\cite{javed2021canintelliids} combined \ac{CNN} with attention-based \ac{GRU} to achieve state-of-the-art performance (99.5\% accuracy) for controller area network intrusion detection.

Hybrid architectures have shown promise for complex vehicular scenarios. Sun et al.~\cite{sun2021anomaly_cnnlstm} proposed a \ac{CNN}-\ac{LSTM} with attention for spatiotemporal feature extraction, demonstrating improved detection of multi-stage attacks. Zhu et al.~\cite{zhu2019mobile_edge_lstm} investigated mobile edge-assisted deployment of \ac{LSTM} models, achieving real-time inference ($<$ 50$ms$ latency) on resource-constrained edge servers. 

\ac{FL} approaches address privacy concerns in distributed vehicular networks. Gurjar et al.~\cite{gurjar2025federated_vanet} proposed a \ac{FL}-based misbehavior classification framework achieving 98.6\% accuracy while preserving vehicle location privacy. However, these approaches typically require substantial computational resources for model training and aggregation, which limits their practical deployment on edge platforms. 

While achieving high accuracy, existing \ac{DL} approaches face limitations: \ac{LSTM} architectures struggle with long-range temporal dependencies essential for analyzing platoon coordination, and \ac{CNN}-based methods require careful manual feature engineering for time series data. Furthermore, most prior work evaluates detection performance in isolation, ignoring an end-to-end pipeline from data collection to edge deployment.

\subsection{Edge AI for Vehicular and IoT Networks}

Edge \ac{AI} has emerged as an enabler for real-time vehicular security, addressing latency and privacy constraints inherent in cloud-centric approaches. Gong et al.~\cite{gong2023edge_intelligence_its} identify three key deployment paradigms: on-vehicle inference, roadside unit processing, and hybrid edge-cloud architectures. Zhang and Letaief~\cite{zhang2020mobile} showed that edge intelligence can reduce end-to-end latency by 10-15x over cloud processing while maintaining comparable accuracy, which is essential for safety-critical vehicular applications with sub-100$ms$ response requirements.

\ac{TinyML} frameworks enable deployment on resource-constrained hardware. De Prado et al.~\cite{deprado2021robustifying} demonstrated deployment of \ac{DNN} models on autonomous mini-vehicles using model quantization, achieving 3-5x energy efficiency improvements with minimal accuracy degradation ($<$ 2\%). Alajlan and Ibrahim~\cite{alajlan2022tinyml} provided a comprehensive overview of \ac{TinyML} techniques for \ac{IoT} edge devices, emphasizing the importance of model architecture and target hardware co-design. 

\ac{DNN} inference acceleration in vehicular edge computing has received substantial research attention. Liu et al.~\cite{liu2023accelerating_dnn} proposed joint optimization of task partitioning and edge resource allocation, achieving 40\% latency reduction with 99\% reliability guarantees for safety-critical inference tasks. Li et al.~\cite{li2025dnn_vec} developed mobility-aware acceleration strategies that adapt model partitioning based on vehicle trajectory predictions, maintaining inference quality during handover events.

\ac{FL} approaches balance privacy preservation with model performance in distributed vehicular environments. Valente et al.~\cite{valente2023embedded_fl} demonstrated embedded \ac{FL} for \ac{VANET} environments, achieving 94.3\% accuracy with on-vehicle training on Raspberry Pi-class hardware. Mughal et al.~\cite{mughal2024adaptive_fl} proposed adaptive \ac{FL} with multi-edge clustering, reducing communication overhead by 60\% while maintaining convergence properties. Pervej et al.~\cite{pervej2023resource_constrained} addressed highly mobile scenarios, developing a resource-constrained vehicular edge \ac{FL} that maintains 93\% accuracy despite frequent topology changes.

Model quantization techniques enable the deployment of \ac{AI} on edge platforms~\cite{2022_jamming_hussain}. Recent work~\cite{hussain2024edge, 2025_lightweight_edge_ai_rff} has demonstrated that 8-bit integer quantization can reduce transformer model sizes by 4x with $<$ 1\% accuracy degradation for classification tasks. However, the impact of quantization on transformer-based misbehavior detection for vehicular platoons remains unexplored, particularly regarding the precision-recall trade-offs critical for safety-critical applications.

\subsection{Positioning of Our Work}
Table~\ref{tab:comparison} compares \textsc{AIMformer} with existing frameworks. Our work builds upon foundational contributions in vehicular misbehavior detection, particularly VeReMi~\cite{van2018veremi} for plausibility-based validation, and extends recent advances in ensemble methods~\cite{ercan2022misbehavior, gyawali2019misbehavior}, \ac{FL}~\cite{uprety2021privacy, gurjar2025federated_vanet}, and recurrent architectures~\cite{alladi2021securing, alladi2023deep, hsieh2021deep_integrated}. While these approaches demonstrate strong detection capabilities in general \ac{VANET} contexts, they face limitations when applied to the strict real-time requirements and dynamic characteristics of platoon systems. 

Within platoon-focused detection, prior approaches span probabilistic methods (GMM-HMM, 87\% F1, 130-280$ms$)~\cite{kalogiannis2022attack}, recurrent networks (LSTM/GRU, $>$ 96\% accuracy)~\cite{wang2021deep, ko2021approach}, and specialized methods (RF, 10$ms$ inference)~\cite{boddupalli2021replace, boddupalli2019redem, mousavinejad2022secure}. Most directly, our work builds on AttentionGuard~\cite{li2025attentionguard}, which achieves 88-92\% accuracy and 96-99\% \ac{AUC} but with inference times of 100-500$ms$, precluding practical edge deployment.

\textsc{AIMformer} advances beyond AttentionGuard and the broader landscape through four contributions: (1) integrated quantization-aware training achieving the first transformer-based detector with sub-millisecond inference (0.13-0.8$ms$ versus 100-500$ms$), enabling practical \ac{OBU}/edge deployment; (2) unified framework handling position, speed, and acceleration falsification including stealth variants through adaptive masking and global positional encoding, whereas prior work targets specific attack types~\cite{boddupalli2021replace} or scenarios~\cite{wang2021deep}; (3) two to three orders of magnitude faster inference while maintaining superior detection performance (96-99\% \ac{AUC}); (4) comprehensive evaluation under maneuvering (join, exit, lane change) representing critical vulnerability windows that prior work does not systematically address~\cite{li2025attentionguard, kalogiannis2022attack, ko2021approach}.

To summarize, \textsc{AIMformer} delivers a complete solution combining high detection accuracy with practical edge deployment feasibility, addressing the fundamental challenge of deploying \ac{DL} models on resource-constrained vehicular hardware while advancing beyond probabilistic~\cite{kalogiannis2022attack}, recurrent~\cite{wang2021deep, ko2021approach}, specialized~\cite{boddupalli2021replace}, and our prior work~\cite{li2025attentionguard}.

%% file: Sections/8_conclusion.tex
\section{Conclusion}
\label{sec:conclusion}
We presented \textsc{AIMformer}, a transformer-based framework for misbehavior detection tailored specifically to vehicular platooning environments. Our framework incorporates domain-specific techniques, including global positioning encoding with temporal offsets, as well as a precision-focused loss function. These components effectively address the unique challenges of detecting malicious behavior in coordinated multi-vehicle systems. 
\rev{\textsc{AIMformer} supports two configurations: $B \cdot V$ captures intra-vehicle temporal dynamics independently per vehicle for distributed vehicle-local deployment, while $V \cdot T$ further captures inter-vehicle spatial correlations through joint spatio-temporal attention for infrastructure-based deployment. Both enable robust detection across heterogeneous control strategies and attack vectors with the computational efficiency required for safety-critical applications.} Our evaluation establishes \textsc{AIMformer}'s superiority over existing \ac{DNN} architectures, demonstrating consistent performance gains across various attack scenarios with an \ac{AUC} of 96-99\%. Moreover, deployment analysis validates the practical feasibility of sub-millisecond inference latencies, achievable through model quantization and optimization, enabling both vehicle-local and infrastructure-based deployments.